\documentclass[aps,prb,preprint,superscriptaddress]{revtex4-2}
\usepackage{graphicx}  
\usepackage{dcolumn}   
\usepackage{bm}        
\usepackage{amssymb}
\usepackage[english]{babel}
\usepackage{booktabs}
\usepackage{array}
\usepackage{caption}
\usepackage{subcaption}
\usepackage{hyperref}
\hypersetup{
	colorlinks=true,
	linkcolor=green,
	filecolor=magenta,      
	urlcolor=blue,
}
\begin{document}

\title{Structural and electronic phase transitions in Zr$_{1.03}$Se$_{2}$ at high pressure}

\author{Bishnupada Ghosh}
\affiliation{National Center for High Pressure Studies, Department of Physical Sciences, Indian Institute of Science Education and Research Kolkata, Mohanpur 741246, Nadia, West Bengal, India.}

\author{Mrinmay Sahu}
\affiliation{National Center for High Pressure Studies, Department of Physical Sciences, Indian Institute of Science Education and Research Kolkata, Mohanpur 741246, Nadia, West Bengal, India.}

\author{Debabrata Samanta}
\affiliation{National Center for High Pressure Studies, Department of Physical Sciences, Indian Institute of Science Education and Research Kolkata, Mohanpur 741246, Nadia, West Bengal, India.}

\author{Pinku Saha}
\altaffiliation[Present address:]{Department of Earth Sciences, ETH Zürich, Zurich 8092, Switzerland}
\affiliation{National Center for High Pressure Studies, Department of Physical Sciences, Indian Institute of Science Education and Research Kolkata, Mohanpur 741246, Nadia, West Bengal, India.}

\author{Anshuman Mondal}
\altaffiliation[Present address:]{Institut für Mineralogie, University of Münster, Münster 48149, Germany}
\affiliation{National Center for High Pressure Studies, Department of Physical Sciences, Indian Institute of Science Education and Research Kolkata, Mohanpur 741246, Nadia, West Bengal, India.}

\author{Goutam Dev Mukherjee}
\affiliation{National Center for High Pressure Studies, Department of Physical Sciences, Indian Institute of Science Education and Research Kolkata, Mohanpur 741246, Nadia, West Bengal, India.}

\email [Corresponding author:]{goutamdev@iiserkol.ac.in}

\begin{abstract}
A detailed high pressure investigation is carried out  using x-ray diffraction, Raman spectroscopy and low temperature resistivity measurements on hexagonal ZrSe$_{2}$ having an excess of 3 at.\% Zr. Structural studies show that the sample goes through a gradual structural transition from hexagonal to monoclinic phase, with a mixed phase in the pressure range 5.9 GPa to 14.8 GPa. Presence of minimum in the $c/a$ ratio in the hexagonal phase and a minimum in the full width half maximum of the $A_{1g}$ mode at about the same pressure indicate to an electronic phase transition. The sample shows a metallic characteristic in its low temperature resistivity data at ambient pressure, which persist till about 5.1 GPa and can be related the presence of slight excess Zr. At and above 7.3 GPa, the sample shows a metal to semiconductor transition with opening of a very small band gap, which increases with pressure. The low temperature resistivity data show an upturn, which flattens with increase in pressure. The phenomenological analysis of the low  temperature resistivity data indicate to the presence of Kondo effect in the sample, which may be due to the excess Zr. 
\end{abstract}

\date{\today}
\maketitle

\section{INTRODUCTION}
The growing importance of transition metal dichalcogenides (TMDs) in opto-electronic materials, data storage devices, sensors, spintronics, etc. has brought them to the focus of current research interest\cite{rehman2018enhanced,song2017anomalous}. For understanding fundamental aspects of strong electron correlations, TMDs are ideal materials due to their diverse physical properties like strong spin orbit coupling, charge density wave, superconductivity and Kondo effect \cite{keum2015bandgap,kvashnin2020coexistence,sahoo2020pressure}. TMDs stabilize in different structures and are associated with different stacking order. These different morphologies lead to distinct electronic behavior, ranging from insulators, semiconductors, semimetal and even superconductors\cite{zhai2018pressure}. Pressure is a very effective physical parameter which can tune the electronic behavior of TMDs by modifying their crystal structures in a controlled way\cite{ghosh2021strain, saha2018structural, saha2020high, saha2020pressure, kumar2015pressure, bandaru2014effect}. Beside these, controlling layer numbers, presence of impurities etc. can tune electronic properties of TMDs significantly\cite{huang2016dynamical, ur2019band, shkvarin2018band, tang2018electronic, wang2021weak}.

Semiconducting TMDs from group IVB (Ti,Zr,Hf) have not been studied so intensively as like their counter parts from  group VIB (Mo,W)\cite{manas2016raman}.
ZrSe$_{2}$ and HfSe$_{2}$ semiconductors have the moderate band gap similar to silicon ranging from 0.9 to 1.2 eV and can be a suitable replacement in the field of electronic devices\cite{salavati2019electronic}. During fabrication of electronic devices, the TMDs are needed to be deposited on substrates, which may produce a large strain within the materials resulting in changes in the structural as well as electronic properties. Therefore high pressure investigation of these materials are extremely important.  A recent theoretical study has shown a pressure induced semiconductor to metal transition associated with a structural phase transition in ZrS$_{2}$ from the hexagonal(H) phase at ambient condition  to the monoclinic phase at 2 GPa followed by the  orthorhombic phase at 5.6 GPa\cite{zhai2018pressure}. A semiconductor to metallic transition is predicted for 1T-ZrTe$_{2}$ and 2H-ZrTe$_{2}$ at pressure of 2 GPa and 6 GPa, respectively\cite{kumar2015pressure}. Also other recent structural investigations in MoTe$_{2}$, WS$_{2}$, MoSe$_{2}$ at high pressure have predicted electronic transitions driven by strain in the unit cell\cite{ghosh2021strain, saha2018structural, saha2020high, saha2020pressure}.

In this work we have carried out a detailed high pressure investigation on hexagonal Zr$_{1.03}$Se$_{2}$ single crystal using x-ray diffraction (XRD) and Raman spectroscopy measurements up to about 34.1 GPa. We have also performed low temperature resistivity ($\rho$)measurements at selected pressure points up to about 8.1 GPa. From XRD study a gradual structural phase transition from hexagonal to monoclinic phase is observed starting from 5.9 GPa, which completes at around 14.8 GPa. The Raman spectra also show noticeable changes above 5.9 GPa, which may be due the change in symmetry associated with the structural transition.  The temperature variation of resistivity of Zr$_{1.03}$Se$_{2}$ reveals a metallic character at ambient pressure due to the presence of excess Zr. At low temperatures, the $\rho$ shows an upturn indicating a possibility of Kondo effect, resulting from the magnetic impurities associated to excess Zr. Above 5.1 GPa the increase in resistivity  with the decrease of temperature shows a metal to semiconductor transition. 

\section{EXPERIMENTAL DETAILS}
High quality single crystal 1T  ZrSe$_{2}$ sample is purchased from HQ graphene (Netherlands)  for carrying out the experiments. XRD characterization of the sample showed it to be in hexagonal crystal structure, which will be discussed in detail in the next section. We have carried out EDAX measurements of the purchased single crystal for determination of chemical composition. The EDAX spectra are shown in supplementary Fig.S1\cite{supplementary}. Taking the average of the data, we find the chemical composition to be Zr:Se=1.03:2 in atomic percentages, indicating the presence of excess Zr in the sample (supplementary Fig.S1\cite{supplementary}).
 
XRD and Raman spectroscopy experiments at high pressures are carried out using piston-cylinder type diamond-anvil cell (DAC) from Almax easyLab. The diameter of culets of the diamond anvils are 300 $\mu$m. A steel gasket is pre-indented  to a thickness of 45 $\mu$m  by applying pressure using the DAC. A hole of 100 $\mu$m diameter is drilled at the center of the preindented gasket. Small ruby chips (approximate size 3-5 $\mu m  $) are loaded along with the sample inside the hole to measure the pressure {\it {in situ}} during Raman spectroscopy measurements\cite{mao1986calibration}. In case of high pressure XRD measurement a very small amount of silver powder is mixed with the sample, which acts as pressure calibrant for {\it {in situ}} pressure measurement.   Methanol-Ethanol mixture in the ratio of 4:1 is used as liquid pressure transmitting medium(PTM) in both the cases to generate hydrostatic pressure. For Raman measurements a Cobolt-samba diode pump laser of wavelength 532 nm is used for excitation. Raman spectra are collected in the backscattering geometry using a confocal micro-Raman system (Monovista from SI GmbH) consisting of a 750mm monochromator and a back-illuminated PIXIS 100BR (1340$\times$100) CCD camera. 20x infinitely corrected long working distance objective with numerical aperture 0.42 is used to collect the scattered light. The scattered light is dispersed using a grating with 1500 grooves/mm having a spectral resolution better than 1.2 cm$^{-1}$.

X-ray source of XPRESS beam-line in ELETTRA synchrotron with the wavelength of 0.4957 \AA\ is used to perform high pressure XRD study of bulk Zr$_{1.03}$Se$_{2}$  using the same diamond anvil cell configuration described above.  The incident X-ray beam is collimated to 20 $\mu m$ diameter using a pin hole and passed through the center of the loaded gasket hole. A MAR 345 image plate detector system aligned normal to the beam is used to collect the diffracted XRD patterns. Sample to detector distance is calculated using the XRD pattern of LaB$_{6}$. Acquired 2D XRD images are converted into 2$\theta$ vs intensity  using Dioptas software\cite{prescher2015dioptas}.
Pressure is calibrated using 3rd order Birch Murnaghan equation of state of silver \cite{dewaele2008compression}.
X-ray diffraction patterns are indexed using CRYSFIRE software\cite{shirley2002crysfire} followed by Rietveld refinements using GSAS software\cite{toby2001expgui}.

To investigate pressure evolution of  electronic behavior of 1T Zr$_{1.03}$Se$_{2}$ single crystal, low temperature and high pressure resistivity measurements are carried out in the range 22-300K, and 0-8.1 GPa respectively. The low temperature and high pressure resistance measurements are performed using a miniature plate type diamond anvil cell with diamonds having culet diameter of 400 $\mu$m. A hole of 200 $\mu$m diameter is drilled in the middle of a steel gasket pre-indented to a thickness of about 30 $\mu$m. A small amount of NaCl is put inside the hole as PTM and prepressed inside the hole. next a weighed mixture of alumina powder and epoxy is spread on the gasket and pressed using two diamond culets by applying a pressure up to about 10 GPa. This helps to insulate the gasket and the gold wires used as probes. A square shaped sample is cut and kept at the middle of the insulated gasket hole. Four thin gold wires pre-pressed and placed on the sample surface, are used as four probes for resistance measurements using the Van der Pauw method, which is well accepted in the resistivity measurements using a DAC\cite{vaughan1961four,philips1958method}. The loaded DAC is placed inside a closed cycle cryostat for low temperature high pressure measurements. Keithley current source (Model No:6221) and nano-voltmeter( Model No:2182A) are used to source the current and measure the voltage, respectively. The schematic diagram of the loaded diamond anvil cell is shown in supplementary Fig.S2\cite{supplementary}. Pressure is measured using the Ruby fluorescence technique. To check the existence of insulation between the gasket and the sample, we carry out a periodic check during the experiments. Measurements are performed by changing the leads alternately and taking the average value of resistivity. In first configuration current is sourced through 1st and 2nd leads and voltage is measured across 3rd and 4th leads, and in second configuration current is sourced through 2nd and 3rd leads and voltage is measured across 1st and 4th leads, to overcome pressure in-homogeneity.
The resistivity of the sample is calculated using the simple formula\cite{gao2005accurate, ghosh2021strain}
\begin{equation}
	\rho=R\frac{A}{l},
\end{equation}
\noindent
where $\rho$ and $R$ are the resistivity and the resistance of sample, respectively; $l$ is the distance between two electrodes; and $A$ represents the product of sample thickness and the lateral width of the sample.
The pressure exerted on the sample is pseudo-hydrostatic. We have used an almost square shaped sample, by exfoliating from a single crystal. Since it was not possible to determine the actual change in thickness of the sample with pressure, we have assumed it to be constant.

\section{RESULTS AND DISCUSSION}

The ambient XRD pattern of Zr$_{1.03}$Se$_{2}$ is indexed to the hexagonal crystal structure  in P$\bar3$m1 space group symmetry with lattice parameters, $a=$3.7661(3) {\AA}, $c=$6.1239(10) {\AA} and unit cell volume $V$=75.219(16) {\AA}$^{3}$ and $Z$=1, which matches well with the literature\cite{manas2016raman}. Rietveld refinement of the ambient XRD pattern is carried out using the atom positions given by Zhai et al.\cite{zhai2018pressure} as the starting model.  Fig.1(a) shows the Rietveld refinement of XRD pattern of Zr$_{1.03}$Se$_{2}$ at ambient pressure. The atomic arrangements inside a unit cell are shown in Fig.1(a). The refined atom positions are given in Table-I. In a layer of Zr$_{1.03}$Se$_{2}$, Zr atoms are sandwiched between Se atoms as shown in Fig.1(a).

In Fig.2, we have presented the pressure evolution of XRD patterns at selected pressure points up to 30.9 GPa (XRD patterns at lower pressure region are shown in supplementary Fig.S3\cite{supplementary}). The XRD Bragg peaks corresponding to the Ag pressure marker are indicated by star. All XRD Bragg peaks broaden with pressure and shift towards higher 2$\theta$ values due to unit cell volume compression. Around 5.9 GPa certain new peaks around 2$\theta$=10.52 deg, 12.47 deg, 12.85 deg, 17.68 deg and 21.10 deg (shown in Fig.2 using arrow) appear. The new XRD pattern could not be indexed to the parent hexagonal crystal structure. Above 5.9 GPa, the intensity of (101) Bragg peak of the parent hexagonal phase reduces continuously and disappears around 14.8 GPa. The intensities of the new peaks, those appeared around 5.9 GPa, grow with pressure. This indicates a gradual growth of a new structural phase with pressure, in parallel with the continuous disappearance of the parent hexagonal 1T phase, indicating a mixed phase in the pressure range 5.9 to 14.8 GPa. The XRD pattern at 14.8 GPa is indexed to a monoclinic structure with a high figure of merit having lattice parameters $a$=4.9785(13){\AA}, $b$=3.1199(4){\AA}, $c$=7.7119(20){\AA}, monoclinic angle $\beta$=91.10(2)${^\circ}$ and unit cell volume $V$=119.76(3) \AA$^{3}$ with $Z$=2. Analyzing our d-values using CHEKCELL\cite{laugier2004chekcell} returned a best space group P2$_{1}$/m. 
We also could index the XRD pattern to an orthorhombic phase in space group $Immm$, similar to that predicted by Martino et al.\cite{martino2020structural}. However, profile fitting followed by the Rietveld refinement using the orthorhombic structure did not result in a good fit (supplementary Fig.S4\cite{supplementary}).
Therefore, we proceed with the Rietveld Refinement of the high pressure phase using the obtained lattice parameters in the monoclinic crystal structure. We used the atom positions of the similar high pressure monoclinic phase of the sister material ZrS$_{2}$ as reported by Zhai et. al\cite{zhai2018pressure}. In Fig.1(b), we have shown the Rietveld refined pattern of our monoclinic phase, which shows an excellent fit. A comparison of the Rietveld refinements between the both high pressure structures are shown in the supplementary information in Fig.S4 and Fig.S5\cite{supplementary}. However, we would like to mention that the Bragg peaks (3 0 -2), (4 0 1), (3 0 2), (2 1 -2) are slightly affected due to specific orientation at high pressure. In Table-I we have given the new atom positions for the monoclinic phase and the unit cell structure is shown in Fig.1(b). The new monoclinic unit cell consists of distorted ZrSe6 octahedra arranged in a layer.

The XRD pattern in the pressure range 5.9-12.1 GPa, now could be fitted using the mixed phase of the parent hexagonal and the new monoclinic phase. With increasing pressure the fraction of the parent hexagonal phase decreases and the new monoclinic phase increases with a complete transformation above 12.1 GPa. The variation of phase fraction obtained from the Rietveld refinements is shown in the supplementary Fig.S6\cite{supplementary}. Rietveld refinement of XRD pattern using mixed phase at 10.4 GPa  is shown in supplementary Fig.S7\cite{supplementary}.
The variation of the relative lattice parameters of hexagonal phase are shown in Fig.3(a). We find that $a$-axis is compressed almost linearly with pressure where as $c$-axis reduces rapidly up to about 5.9 GPa with a compression of about 5.7\%, and then it almost saturates in the mixed phase. To understand the anomalous compression behavior of the hexagonal unit cell, we have plotted $c/a$ ratio in Fig.3(b), which shows a sharp minimum at about 5.9 GPa. The $c/a$ ratio is associated with the lattice strain in the hexagonal phase. Decreasing $c/a$
ratio indicates a decrease in internal strain of the unit cell till the phase transition to the mixed phase. In layered materials, due to Van der Waals interaction among the layers along $c$-axis, initial increase in pressure is generally found to compress the $c$-axis significantly in comparison to the $ab$ plane. However, above 5.9 GPa, in the compact structure, the $c$-axis can not be compressed anymore significantly, resulting in an increase in the lattice strain in the $ab$-plane. It may be suggested that the anisotropic compression of the parent hexagonal phase leads to such a structural transition, which results in increase in the lattice strain in the compressed unit cell of the parent structure.
The pressure evolution of relative lattice parameters of monoclinic phase from 5.9 GPa to 30.9 GPa are shown in Fig.3(c). The $b$- and $c$-axes are found to reduce almost linearly with pressure for the whole pressure region. The $a$-axis decreases with pressure up to 12.1 GPa then it starts to increase at higher pressures. The monoclinic angle $\beta$ also shows a similar variation. It decreases with pressure at first but then starts to increase above 17 GPa. At highest pressure of this experiment $a$-axis increases by 8.34 $\%$ whereas $b$- and $c$-axes decrease by 9.3 $\%$ and 14.16 $\%$ of their initial values, respectively . The $c$-axis is the most compressible axis. The pressure evolution of relative lattice parameters shows an anisotropic compression along different directions resulting in differential strain inside the sample. At higher pressures, where $b$- and $c$-axes experience compressive strain, tensile strain develops along the $a$-axis.

The variation of unit cell volume with applied pressure for both the phases are shown in Fig.3(d). The unit cell volume vs pressure data for hexagonal phase is fitted using single 3rd order BM EOS\cite{birch1947finite, ahmad2012theoretical}. The Bulk modulus $B_{0}$ and its first pressure derivative $B^\prime$ for hexagonal phase are found to be 29$\pm$2 GPa and 9$\pm$1 respectively, indicating a large pressure dependent compressibility.
The unit cell volume of the monoclinic phase shows a slope change at the boundary of mixed-phase and the pure monoclinic phase of the sample with a possibility of an isostructural transition. We have not carried out any EOS fitting in the small pressure range of 5.9 to 14.8 GPa due to less number of data points. 
Pressure vs unit cell volume of monoclinic cell above 14.8 GPa is fitted using a 3rd order BM EOS. This gives a value of V$_{0}$=133.1 \AA$^{3}$ and B$_{0}$ =115$\pm$7 GPa and $B^\prime$=4.1$\pm$0.7. The monoclinic phase has a bulk modulus 4 times larger than the hexagonal phase. Extending the high pressure EOS fit of the monoclinic phase down to about 5.9 GPa (black line in Fig,3(b)) shows that it represents well the pressure behaviour of volume up to about 10 GPa. The monoclinic phase volume seems to be more compressible in the range 5.9 to 10 GPa in comparison to that above 10 GPa.  Since the liquid PTM freezes above 10 GPa, the effect of non-hydrostatic stress resulting in a differential strain on the layered material can not be ruled out.

Ambient Raman spectrum is shown in Fig.4(a). We observe an intense mode at 195 cm$^{-1}$ and a mode with lower intensity at 147 cm$^{-1}$ . In addition we observe a very low intense peak at 109 cm$^{-1}$ and a broad hump in the region from 235 cm$^{-1}$ to 280 cm$^{-1}$. The hump like feature can be fitted into two Raman modes at around 239 cm$^{-1}$ and 267 cm$^{-1}$ as shown in the inset of Fig.4(a). The spectrum is found to be similar to that reported by Manas-Valero et al.\cite{manas2016raman}. Following the literature, we have indexed the Raman modes as follows: (i) 195 cm$^{-1}$ as A$_{1g}$, (ii) 147 cm$^{-1}$ as E$_{g}$ (iii) 109 cm$^{-1}$, 239 cm$^{-1}$ and 267 cm$^{-1}$ as two phonon modes. A$_{1g}$ mode is due to the out of plane vibration of Se atoms against each other and E$_{g}$ mode is due to the in-plane vibrations of Se atoms against each other\cite{manas2016raman, nikonov2018synthesis}.
In Fig.4(b) pressure evolution of Raman spectra at selected pressure points are shown up to 34.1 GPa (Pressure evolution of all Raman spectra are shown in supplementary Fig.S8\cite{supplementary}). The mode frequency of A$_{1g}$ mode shifts towards higher wave number with pressure, and its intensity decreases with increasing pressure.  
At 5.9 GPa, changes in the shape of the Raman spectrum are observed in the form of appearance of broad humps around the parent E$_{g}$ and A$_{1g}$ modes. These broad humps give rise to well defined peaks above 7.2 GPa. These new peaks are labeled as P1, P2, P3, P4, P5 at around 108 , 127, 166, 245 and 272 cm$^{-1}$, respectively and are marked in Fig.4(b) at 7.9 GPa. 
The appearance of new Raman modes can be attributed to the change in symmetry in the crystal around 5.9 GPa due to the appearance of new monoclinic crystal structure along with parent hexagonal crystal structure. The XRD results show a mixed phase in between 5.9-12.1 GPa. Possibly, the Raman modes due to both the phases overlap and show a broad feature in the Raman-spectra as seen at 5.9 GPa and 6.5 GPa (Fig.4(b)). The pressure evolution of Raman shift for all the detected modes are shown in Fig.5(a). The pressure variation of A$_{1g}$ mode frequency is fitted linearly with a slope of 3.61 cm$^{-1}$GPa$^{-1}$ up to 6.5 GPa and then almost saturates above 7.2 GPa. Even though the compressibility in the monoclinic phase is much lower compared to the hexagonal phase, it can't justify the sudden saturation-effect of the $A_{1g}$ mode alone, since all other modes show a reasonable pressure effect. 
The change in FWHM with pressure for the A$_{1g}$ mode is shown in Fig.5(b). FWHM of the A$_{1g}$ mode shows a minimum around 5 GPa(shown in inset of Fig.5(b)) and then suddenly jumps above 6.5 GPa  pressure. A decrease in the Raman mode bandwidth is associated with an increase in phonon life time due to decrease in anhramonic scattering. The minimum in the FWHM of the Raman mode is generally associated with an electronic transition in these layered materials\cite{saha2018structural, saha2020high}. It may be noted here that $c/a$ ratio also shows a minimum at the same pressure.
The hydrostatic limit of the methanol-ethanol PTM is  10 GPa\cite{klotz2009hydrostatic}, which is much higher than the pressure where FWHM of A$_{1g}$ mode starts to increase rapidly, indicating the observed effect is due to pressure induced changes in the sample physical properties. The sudden jump above 6.5 GPa can be associated with the increased anharmonic phonon-phonon interactions due to the structural transition to the monoclinic phase along with the increase in lattice strain. 

In order to study the electronic behavior of this material under pressure, we have carried out resistivity($\rho$) {\it {vs}} temperature(T) measurement at several pressure values. The temperature variation of resistivity from room temperature to 22 K at several pressure values is shown in Fig.6. At all pressure values till about 5.1 GPa, the resistivity is found to decrease with decrease in temperature indicating a metallic characteristic and then show an upturn at much low temperatures.
At ambient pressure the resistivity curve takes an upturn at about 58 K and increases with the decrease in temperature showing a broad minimum.
We find that the initial slope of the decrease in $\rho$ vs T decreases with pressure and also the low temperature minimum becomes more and more flat. At 7.3 GPa, we find a rise in $\rho$- values with decrease in T, and also the  $\rho$-value obtained at lowest temperature shows an large increase. Another important aspect is that the resistivity also decreases with pressure at room temperature up to about 5 GPa, followed by a saturation, as shown in inset of Fig.6. 

In several previous theoretical studies  ZrSe$_{2}$ has been assigned as an indirect band gap semiconductor having a moderate band gap in the range of 0.9-1.2 eV similar to silicon\cite{martino2020structural, ghafari2018opposite}. A recent study on a sister TMD, ZrTe$_2$ has shown it to be metallic and EDAX measurements have shown an excess of about 10\% Zr in the host\cite{wang2021weak}. Intercalation of trace amount of Re (about 1.3\% wt.) inside the layers of pure ZrSe$_2$ has resulted in metallic behaviour of the sample\cite{ur2019band}. Another theoretical study has shown that structural defects in the pristine ZrSe$_2$ crystals can change its magnetic and electronic properties including metallic behaviour\cite{gao2018electronic}. Therefore, we believe that the presence of additional 3\% of Zr may result in the metallic characteristic shown by the present sample at ambient and high pressures.
Let us first look into the temperature dependence of resistivity at ambient pressure. In our experiments it is found that the resistivity varies linearly with temperature above 200K. Variation of resistivity linearly with temperature is a signature of electron-phonon scattering because the total number of phonons for scattering at high temperature is directly proportional to T\cite{ashcroft1976solid}. This signifies phonon scattering is the dominant scattering mechanism at high temperature. Above 195K the $\rho$(T) vs T data is fitted to the expression:
\begin{equation}
 \rho(T) = C + DT,
\end{equation}
  where $C$ and $D$ are the fitting parameters. Below 195K up to 70K, resistivity doesn't vary linearly with temperature. In this region $\rho$ vs T data is best fitted to the expression:
\begin{equation}
 \rho(T) = \rho_{0}+ AT^{2} + BT^{5},
\end{equation}
\noindent where $\rho_{0}$, $A$ and $B$ are the fitting parameters. At low temperature, the electron-phonon scattering has a T$^{5}$ dependence\cite{ashcroft1976solid}. The T$^{2}$ dependence of resistivity comes from the electron-electron scattering\cite{thompson1975electron}. Below 58K, the $\rho$ is found to increase with lowering in T, indicating a possibility of a metal to insulator transition. Therefore we have tried to fit our data to $ln\rho$ vs 1/T, which did not yield a linear nature(Supplementary Fig.S9\cite{supplementary}) and hence one can discard a metal to insulator transition in the sample at these low temperatures. 

In strongly correlated layered systems, the upturn in $\rho(T)$ behaviour can be attributed to the electron-electron interaction, weak localization or Kondo effect\cite{xu2006low, ziese2003searching, okuda1999low, lee1985disordered}. In the absence of high pressure resistivity measurements under magnetic field, we shall try to use the phenomenological models and discuss the possible phenomena for the upturn in resistivity at low temperatures. The application of pressure reduces the resistivity of the sample as well as the upturn becomes flatter till about 5.1 GPa. Therefore due to increase in conductivity with pressure, one can ignore the effect of weak localization. Similarly, the electron - electron interaction gives a T$^{-\frac{1}{2}}$ dependence in $\rho(T)$\cite{xu2006low, okuda1999low}. However, as shown in the Supplementary Fig.S10, we do not see any linear dependence of ln$\rho(T)$ with respect to $lnT$.
The electron configuration of Zr atom is [Kr]5s$^{2}$4d$^{2}$. The presence of two unpaired electrons at 4d orbital gives rise to a spin magnetic moment S=1. Therefore, the presence of excess Zr atoms in our sample can act as magnetic impurity embedded in the material. The magnetic impurity can give rise to the resistivity upturn due to the Kondo effect. A recent work on a sister layered TMD, ZrTe$_2$, an upturn in the resistivity values at low temperatures  is reported and is explained due to the Kondo effect arising from additional Zr atom present in the material \cite{wang2021weak}. Therefore, in the absence of high pressure low temperature resistivity measurements under magnetic field we have treated our low temperature $\rho$ vs T data using phenomenological Kondo model. Following the literature, we have fitted our resistivity data at low temperatures, to the equation \cite{wang2021weak,hamann1967new}:
\begin{equation}
	\rho(T)=\rho_{C}+AT^{2}+BT^{5}+\rho_{K}\left ( 1-\ln \left ( \frac{T}{T_{K}} \right )\left \{ \ln ^{2}\left ( \frac{T}{T_{K}} \right )+S(S+1)\pi^{2} \right \}^{-\frac{1}{2}} \right ),
\end{equation}
\noindent where the fitting parameters $\rho_{C}$, $\rho_{K}$, $T_{K}$ and S represent the residual resistivity, temperature-independent resistivity, Kondo temperature and the average spin of magnetic impurities,
respectively. The values of $ A $ and $ B $ are extracted from fitting of region II(195K to 70K) to the Eq.(3). The data of the region III(101K to 22K) is fitted to the Eq.(4) by keeping the values of parameters A and  B fixed to the values obtained from the earlier fit. All the parameter values are given in Table II.
From the fit of temperature dependent resistivity data at ambient pressure, the average impurity spin per Zr atom is obtained to be S= 0.0326. Then the spin for excess Zr can be calculated as (0.0326$\times$1.033)$\div$0.033=1.02$\approx$1. This is consistent with the expected total spin value of Zr atom having electron configuration [Kr]5s$^{2}$4d$^{2}$. The $\rho$ vs T data at high pressure up to 5.1 GPa are similarly analyzed in three different temperature regions(supplementary Fig.S11\cite{supplementary}). The depth of the Kondo anomaly reduces with the application of pressure. The S values seems to remain almost constant till 5.1 GPa(Table II). However, T$_{K}$ is found to increase with pressure.
At and above 7.3 GPa the nature of temperature variation of resistivity changes dramatically. It starts to increase with the lowering of temperature showing a semiconducting type behavior (Fig.7(b)) at 7.3 GPa. The observation of a pressure induced metal to semiconductor transition is rarely reported in literature. In a semiconductor the resistivity varies with temperature as:
\begin{equation}
	\rho = \rho(0) exp^{\left (\frac{E_{g}}{2K_{B}T}  \right )},
\end{equation}
where K$_{B}$ is the Boltzmann constant and E$_{g}$ is the band gap.
But the data at 7.3 and 8.1 GPa could not be fitted using only exponential term as the $ln \rho$ vs 1/T does not show a linear behaviour (Supplementary Fig.S12, Fig.S13\cite{supplementary}). We have also tried to fit the data to the Eq.(4), which also did not produce a good fit (Supplementary Fig.S13\cite{supplementary}). We would like to point out here that our experimental set up limits the low temperature value to about 22 K at the sample. Therefore, we carried  out a linear fit in the range of 22 - 35 K for the data at 7.3 GPa and 22 - 50 K for the data at 8.1 GPa for $ln \rho$ vs 1/T. This produced a $E_g$ value of about 1.5(1) meV and 4.4(3) meV at 7.3 GPa and 8.1 GPa, respectively. Such small values of bandgap indicates that at these high pressures, the $Zr_{1.03}Se_2$ is an almost zero gap semiconductor. Matsuoka and Shimizu reported a similar semimetal to zero gap semiconductor transition in Li at extreme high pressures, where the band gap was estimated to be about 0.83 meV at about 78 GPa \cite{matsuoka2009shimuzu}. But in the present case the value of $E_g$ increased with pressure, where it was found to decrease with pressure for Li. When we fitted the combined Eq.(4) and Eq.(5) to our data by keeping the value of $E_g$ constant, it produced a good fit (Fig. 7(b)), but with larger values of $T_K$ and $S$ in comparison to the values obtained from pure hexagonal phase. Also the mixed phase open up a very small band gap.

In a theoretical study, Zhai et. al\cite{zhai2018pressure} have shown the monoclinic phase of ZrS$_{2}$ to be a narrow band gap semiconductor with band gap of 0.738 eV. It has been observed in several hcp metals and alloys, that the $c/a$ ratio affects the electronic orbital hybridization and can result in change in the electronic band structure near the Fermi surface. The observed minimum in $c/a$ ratio in the present case may be driving a change in the electronic band structure and may result in opening of a small gap near Fermi surface\cite{haussermann2001origin, fujinaga1993effect, sinha2019evolving}.
In fact a recent study has shown that tuning the strain on a monolayer 1T$^\prime$ WTe$_2$ has led to a semimetal to a topological insulator transition \cite{zhao2020strain}. Our XRD studies show that at high pressures beyond 5.9 GPa, both the hexagonal and the monoclinic phases show anisotropic compression behaviour along different axial directions, resulting in differential strain in the unit cell. Therefore, it may also cause a change in the Fermi surface topology resulting in opening of a very small band gap. 

Other systematic behaviour those observed, are, a decrease in the ambient temperature $\rho$ till about the phase transition pressure, slight increase in the average impurity spin and $T_K$ with pressure. The hexagonal phase shows a large compressibility resulting in a large decrease in volume, which includes about 6.4\% decrease in the average Zr-Se bond length and a reduction in the unit cell strain. This structural change possibly results in further enhancement of overlap of electronic orbitals leading to decrease in the electron scattering events and hence increase in electronic conduction till about 5.1 GPa. Investigations on the pressure effect of Kondo temperature in several systems have shown an increase in Kondo temperature with pressure\cite{schilling1973effect, maple1971low, kim1970kondo}. In Cu:Fe and Au:Fe Kondo alloys the pressure induced doubling of Kondo temperature has been explained using a simple model of pressure-induced broadening of the bound state of the impurity Fe-atom and hence increasing the intermixing of the bands \cite{schilling1973effect, maple1971low, kim1970kondo}. In another study on the Ce and Yb based heavy fermion compounds, it has been shown that the increase in hybridization of f and conduction electron states result in increase of Kondo temperature with pressure\cite{goltsev2005origin}. Therefore, in view of the above facts we believe that the increase in intermixing of the d-electron states of the Zr atom resulting in broadening of the bound state of the impurity atom due to large decrease in volume and anisotropic compression of the lattice leads to a large change in $J_{exchange}$ value in the lattice. In fact the XRD studies show that there is an increase in the effective coordination number of $Zr-Se6$ octahedra from 1 in the hexagonal structure to a value about 1.17 in the monoclinic structure at about 7 GPa, indicating modification of $J_{exchange}$ value leading to an increase in the Kondo temperature and the average spin. But additional theoretical and high pressure experimental studies under magnetic field are necessary to get to a definite conclusion of the observed interesting behaviour in the present work. However, the present work shows a close relationship between the effect of impurity and strain on the electronic behaviour of transition metal dichalcogenides.

\section{CONCLUSIONS}
Hexagonal Zr$_{1.03}$Se$_{2}$ undergoes a pressure induced gradual structural phase transition from hexagonal 1T phase to monoclinic 1T$^{'}$ phase which starts around 5.9 GPa  and completes around 14.8 GPa. High pressure Raman measurements show the appearance of several new Raman modes above 5.9 GPa, similar to the pressure where new structural phase starts to grow along with the parent phase. The appearance of new Raman modes is attributed to the change in symmetry in the sample due to the transition to a lower symmetry monoclinic phase. A minimum in FWHM of A$_{1g}$ Raman mode at about 5 GPa shows an increase in phonon life time and indicate to the presence of an electronic phase transition.
The electronic phase transition is confirmed by the low temperature and high pressure resistivity measurements of the sample. The low temperature resistivity data at all pressure values show an upturn which is possibly due to Kondo anomaly in the sample due to the presence of additional Zr which acts as the magnetic impurity. With application of pressure the depth of the Kondo anomaly reduces. The temperature dependent resistivity data show that in the mixed phase a small band gap opens in the sample resulting into a pressure induced metal to semiconductor transition.  
\newline
\newline
{\bf ACKNOWLEDGMENTS}
\newline
The authors gratefully acknowledge the Ministry of Earth Sciences, Government of India, for the financial support under the grant No. MoES/16/25/10-RDEAS to carry out this high pressure research work. B.Ghosh also gratefully acknowledge Department of Science and Technology, Government of India for their INSPIRE fellowship grant for pursuing Ph.D. program. The authors also gratefully acknowledge the financial support from the Department of Science and Technology, Government of India to visit XPRESS beamline in the ELETTRA Synchrotron light source under the Indo-Italian Executive Programme of Scientific and Technological Cooperation, and kind support from the beam line scientist associated with EXPRESS beam line.

\newpage
\begin{table}
	\caption{\label{tab:I}Structural parameters of different phases}
	\begin{ruledtabular}
	
		\begin{tabular}{lllllll}
			\textbf{Phase} & \textbf{Lattice Parameters} (\AA) & \textbf{Atoms} & \textbf{x} & \textbf{y} & \textbf{z} & \textbf{Pressure (GPa)}\\
			\hline
	     	P$\bar3$m1 & $a$=$b$=3.7661 (3) & Zr1(1b) & 0(0)& 0(0)& 0.5000(0)& 0.0\\
			           & $c$= 6.1239 (10) & Se1(2d) & 0.6667(0) & 0.3333(0) & 0.7852(3) & \\\\
			P2$_{1}$/m & $a$= 4.9785 (13) & Zr1(2e) & 0.6686(4) & 0.2500(0) & 0.3694(8) & 14.8\\
			           & $b$= 3.1199 (4) & Se1(2e) & 0.8470(2) & 0.2500(0) & 0.5788(2) & \\
			           & $c$= 7.7119 (20) & Se2(2e) & 0.8220(3) & 0.2500(0) & 0.0473(2) & \\
			           &$\beta$=91.10 (2) & & & & & 
		\end{tabular}
	\end{ruledtabular}
\end{table}

	\begin{table}
	\caption{\label{tab:II} Fitting parameters obtained from resistivity studies at different Pressure points}
	\begin{ruledtabular}
		
		\begin{tabular}{llllllll}
			\textbf{Pressure } & \textbf{$\rho_{c}$$\times$10$^{-6}$ } & \textbf{A}$\times$10$^{-11}$ & \textbf{B}$\times$10$^{-18}$ & \textbf{$\rho_{k}$ }$\times$10$^{-7}$ & \textbf{T$_{k}$ } & \textbf{S}& \textbf{D}$\times$10$^{-8}$\\
			(GPa)&($\Omega$-m)&($\Omega$-m.K$^{-2}$)&($\Omega$-m.K$^{-5}$)&($\Omega$-m )&(K)&&$(\Omega$-m.K$^{-1}$ )\\
			\hline
			0 & 20.79(6) & 35.4(8) & 11.6(9) & 349(6) & 12.3(8) & 0.032(4)& 21.9(1)\\
			0.1 & 10.89(4) & 36(1) & 13.8(2) & 23(2) & 22(1) & 0.048(5) & 10.6(1)\\
			1.5 & 5.677(4) & 15.1(1) & 6.6(2) & 7.2(2) & 24.5(4) & 0.034(1)&6.74(3)\\
			2.8 & 4.261(5) & 5.81(4) & 5.2(7) & 3.0(2) & 25.0(8) & 0.030(2)&	4.15(1)\\
			5.1 & 2.891(7) & 3.57(5) & 1.8(1) & 0.8(1) & 31(2) & 0.03(1)&1.90(2)\\
				
		\end{tabular}
	\end{ruledtabular}

\begin{ruledtabular}
	
	\begin{tabular}{llllll}
		\textbf{Pressure} & \textbf{$\rho_{0}$ } & \textbf{E$_{g}$ }  & \textbf{$\rho_{k}$ } & \textbf{T$_{k}$ } & \textbf{S}\\
		(GPa)&($\Omega$-m $\times$10$^{-6}$)& (eV)&($\Omega$-m $\times$10$^{-5}$)&(K)&\\
		\hline
		7.3 & 19(6)  &  0.0015(1) & 1.57(5) & 60(1)& 0.16(1)\\
		8.1 & 20(1)  & 0.0044(3) & 3.21(7) & 51(1) & 0.13(1)\\
		
	\end{tabular}
\end{ruledtabular}
\end{table}

\begin{figure}
	\centering
	\includegraphics[width=\columnwidth]{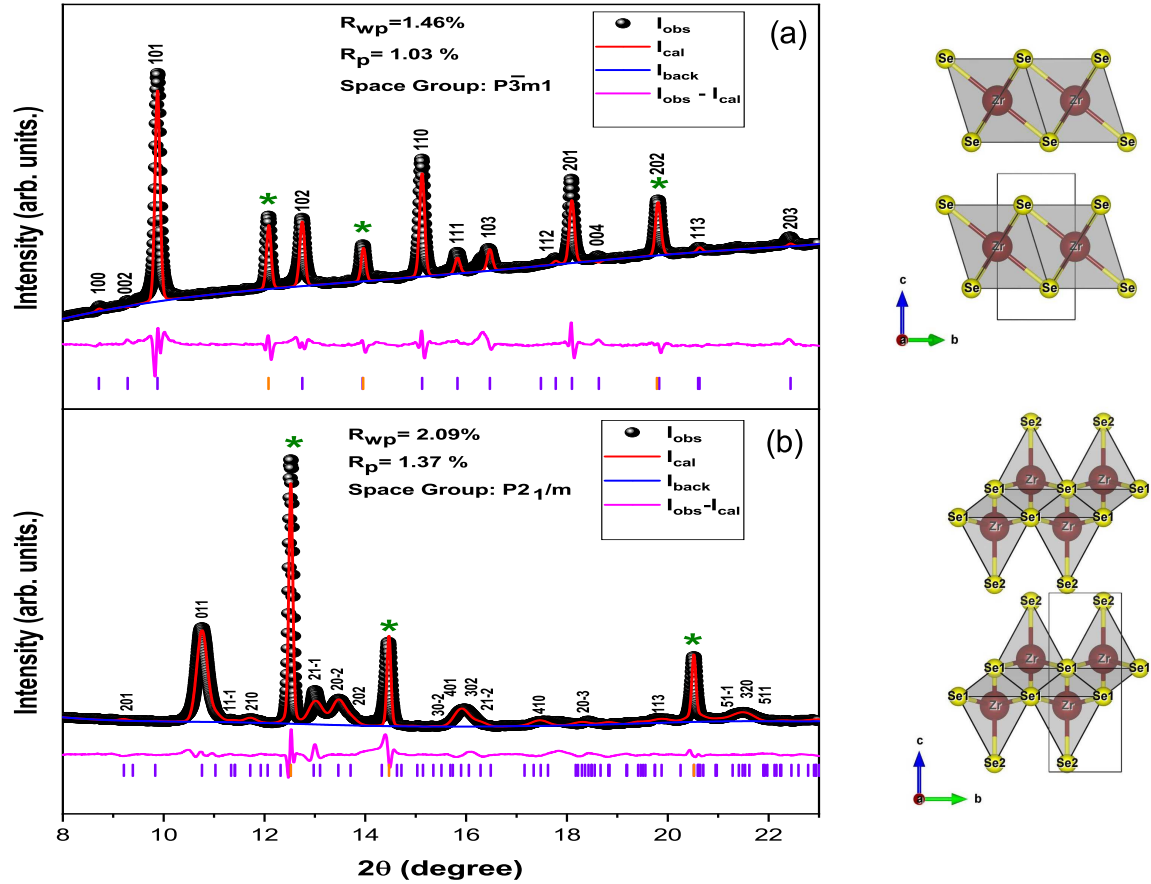}
	\caption{\label{Fig.1} Rietveld refinements of (a) XRD pattern at ambient pressure in the hexagonal phase,  (b) XRD pattern at 14.8 GPa pressure in the monoclinic phase. Atomic orientations inside unit cell of Zr$_{1.03}$Se$_{2}$ crystal in hexagonal as well as monoclinic phase are shown in the right side of the graphs. Silver peaks are marked by green star.}
\end{figure} 

\begin{figure}
	\centering
	\includegraphics[width=\columnwidth, height=20cm]{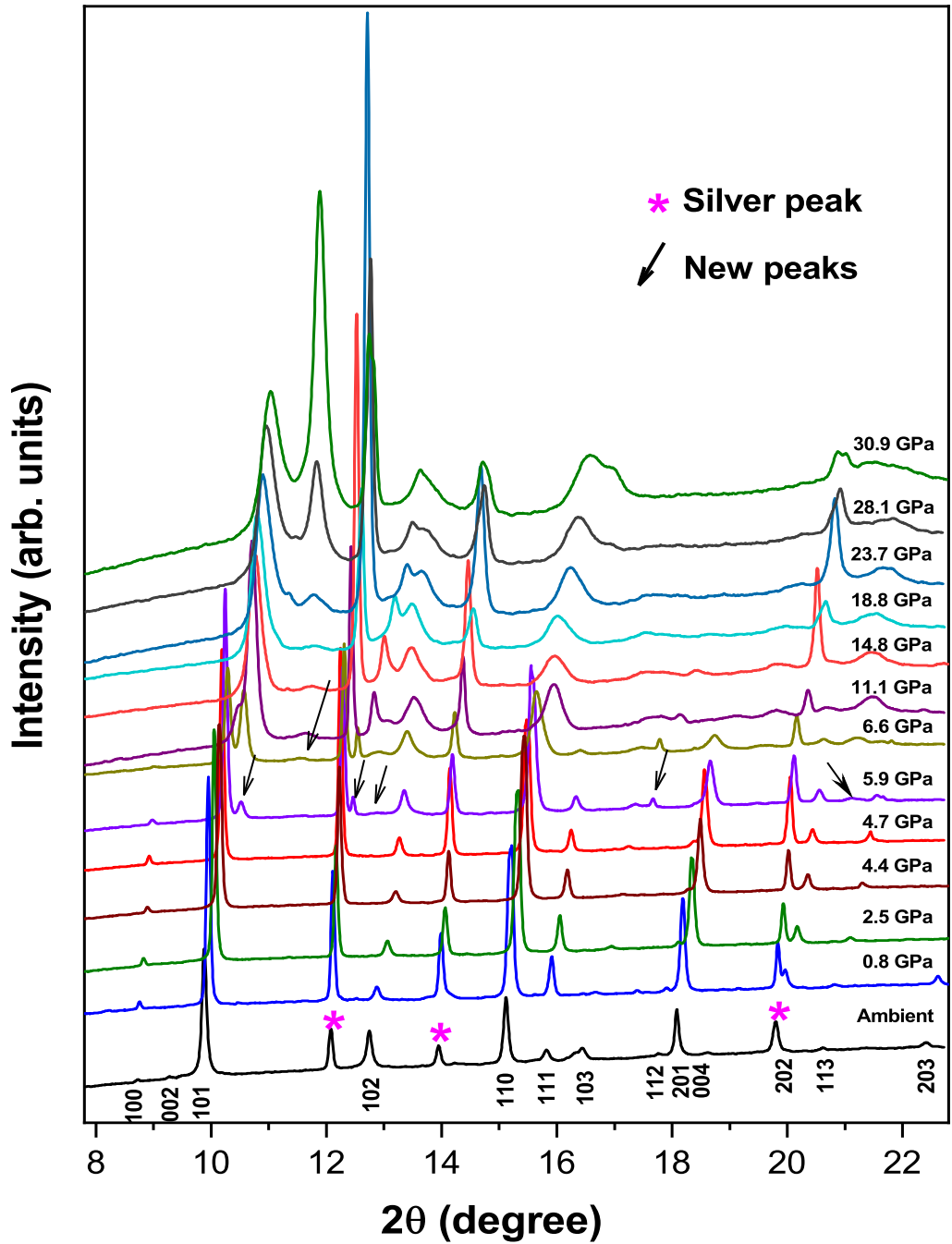}
	\caption{\label{Fig.2}Pressure evolution of XRD patterns. Ag peaks are shown with pink star. New Bragg peaks appearing around 5.9 GPa are denoted by arrows.}
\end{figure}

\begin{figure}
	\begin{center}
		\includegraphics[width=\columnwidth,height=20cm]{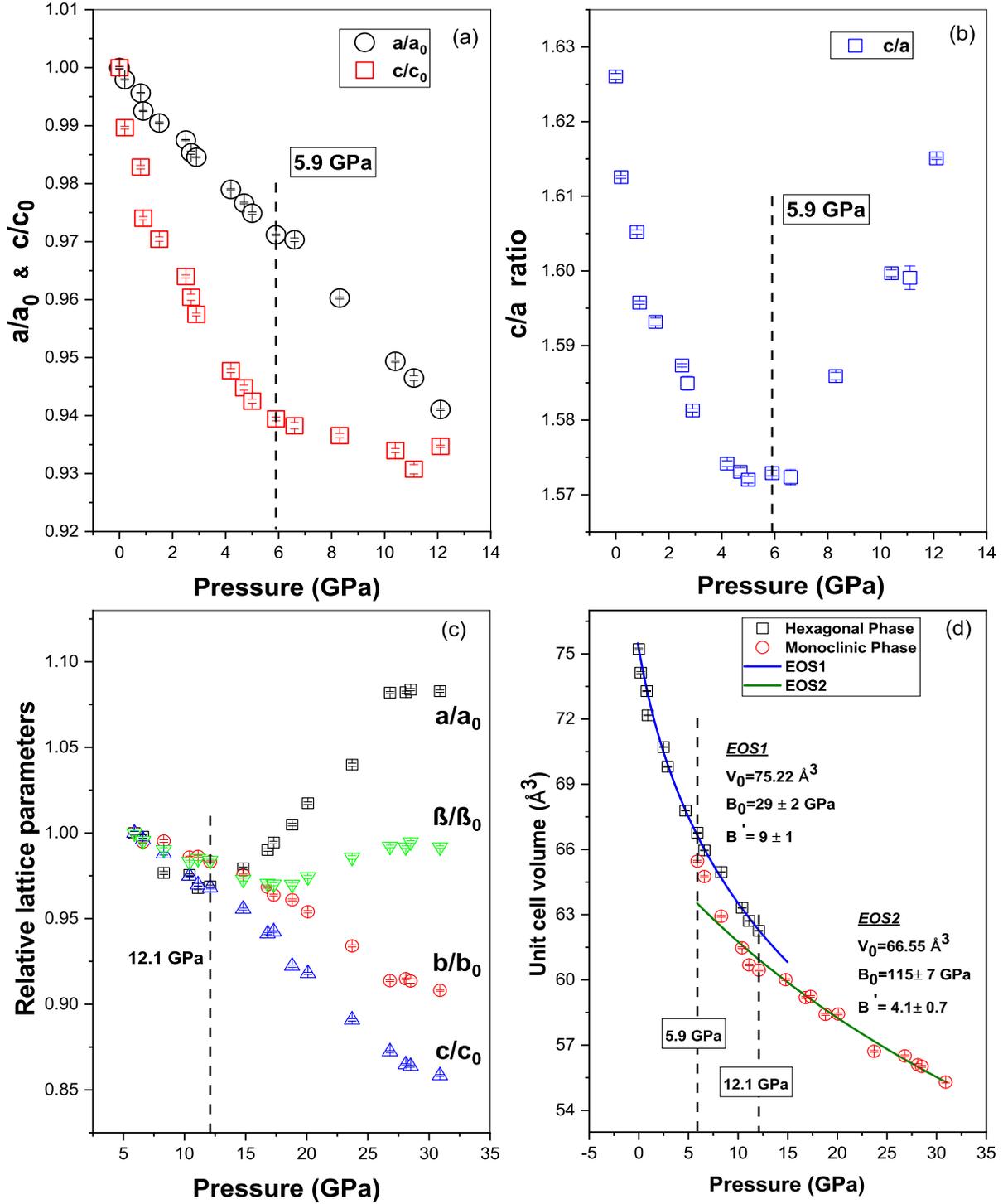}
	\end{center}
	\caption{\label{Fig.3} Pressure evolution of (a) the relative lattice parameter of hexagonal phase, (b) c/a ratio of hexagonal phase and (c) relative lattice parameters of monoclinic phase. (d) The variation of unit cell volume with applied pressure for two phases. The lines through data points show the EOS fits.}
\end{figure}

\begin{figure}
	\centering
	\includegraphics[width=\columnwidth,height=20cm]{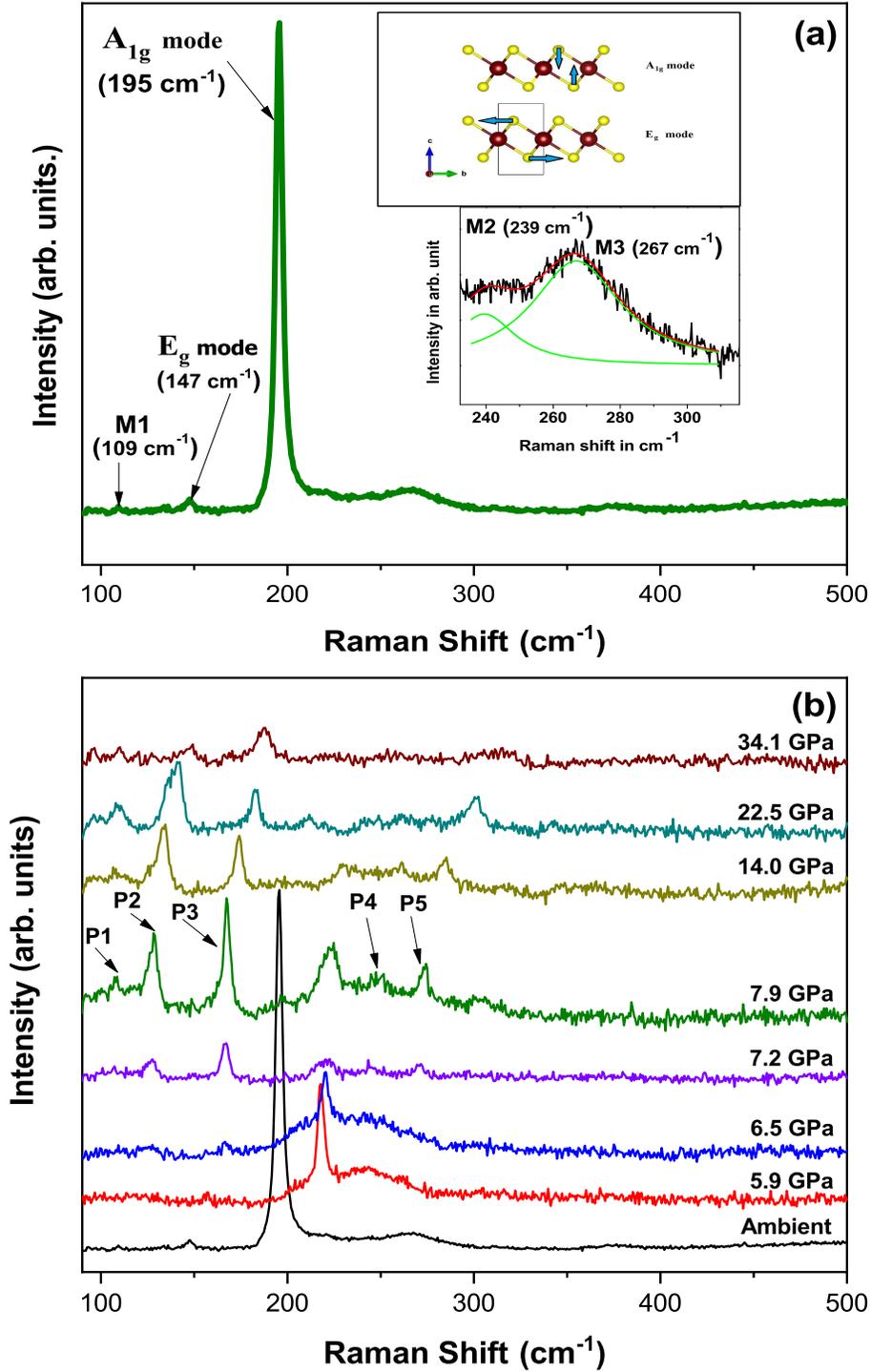}
	\caption{\label{Fig.4} (a) Ambient Raman spectrum of Zr$_{1.03}$Se$_{2}$ using 532nm laser as excitation source. E$_{g}$ and A$_{1g}$ mode vibrations are shown in the inset. Two phonon modes around 239 cm$^{-1}$ and 267$^{-1}$ are also shown in the inset of the figure. (b) Pressure evolution of Raman spectra at selected pressure points. P1,P2,P3,P4,P5 denote the new Raman modes in the monoclinic phase.}
\end{figure}

\begin{figure}
	\centering
	\includegraphics[width=\columnwidth]{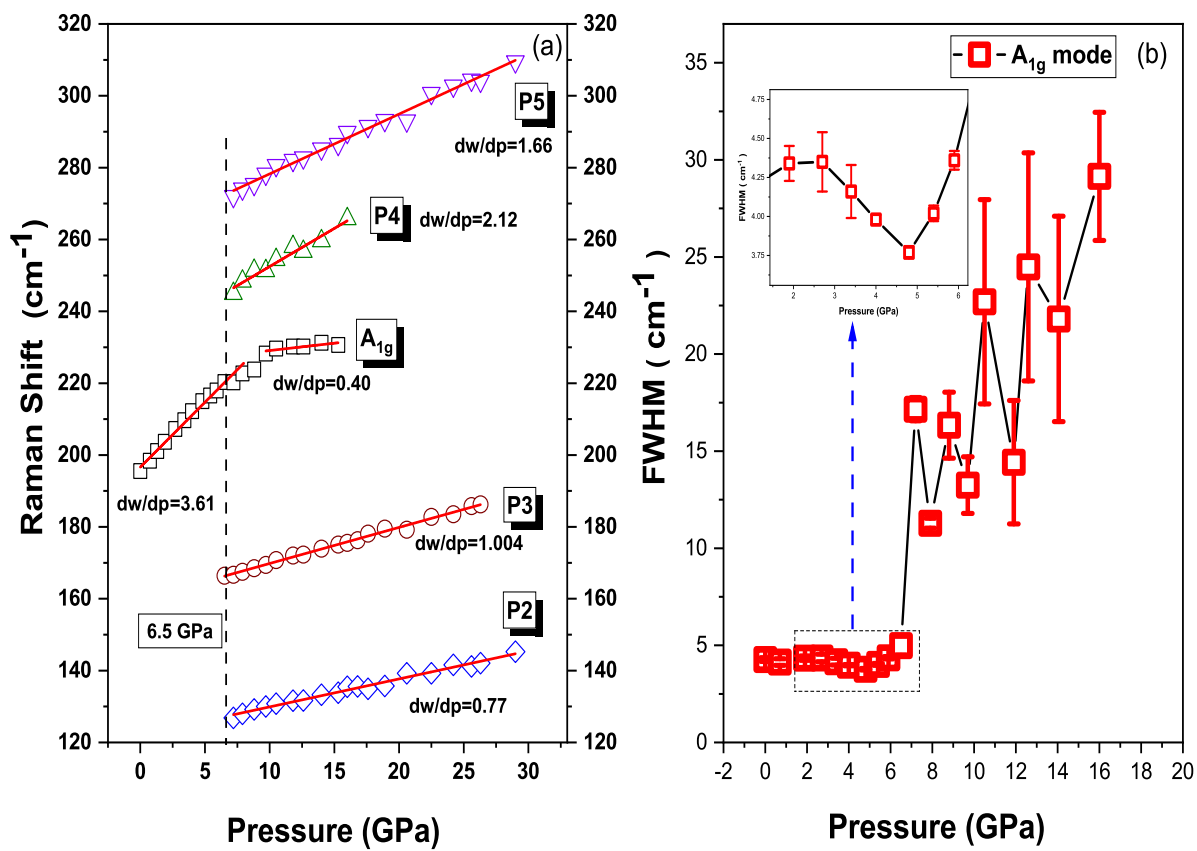}
	\caption{\label{Fig.5}(a)Pressure variation of Raman mode frequencies. (b) Pressure variation of FWHM of A$_{1g}$ Raman mode, with the inset showing the zoomed part marked by the box.}
\end{figure}

\begin{figure}
	\centering
	\includegraphics[width=\columnwidth]{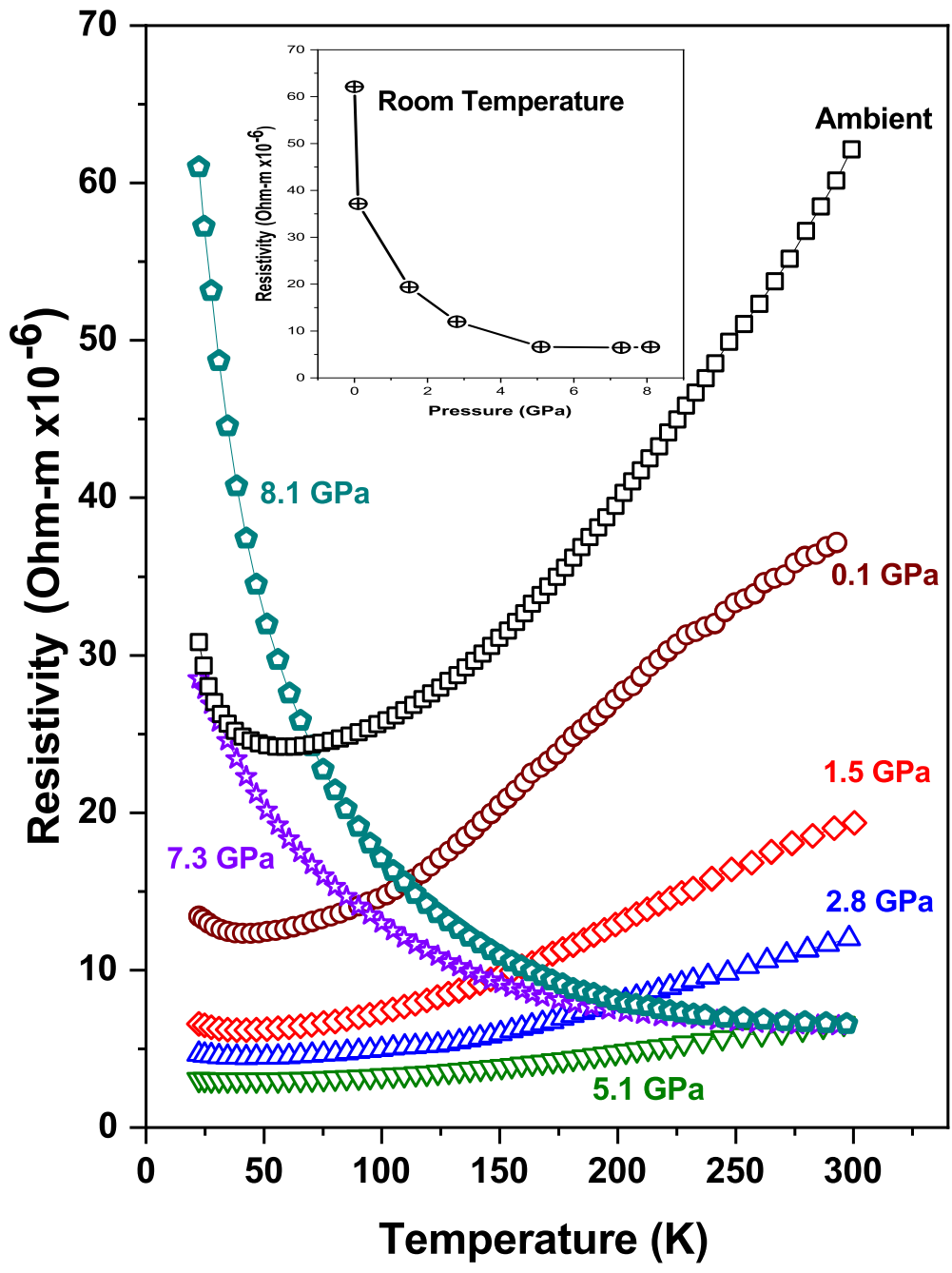}
	\caption{\label{Fig.6} Temperature variation of resistivity at different pressure points. Pressure variation of Room temperature resistivity is shown in inset.}
\end{figure}

\begin{figure}
	\centering
	\includegraphics[width=15cm, height=20cm]{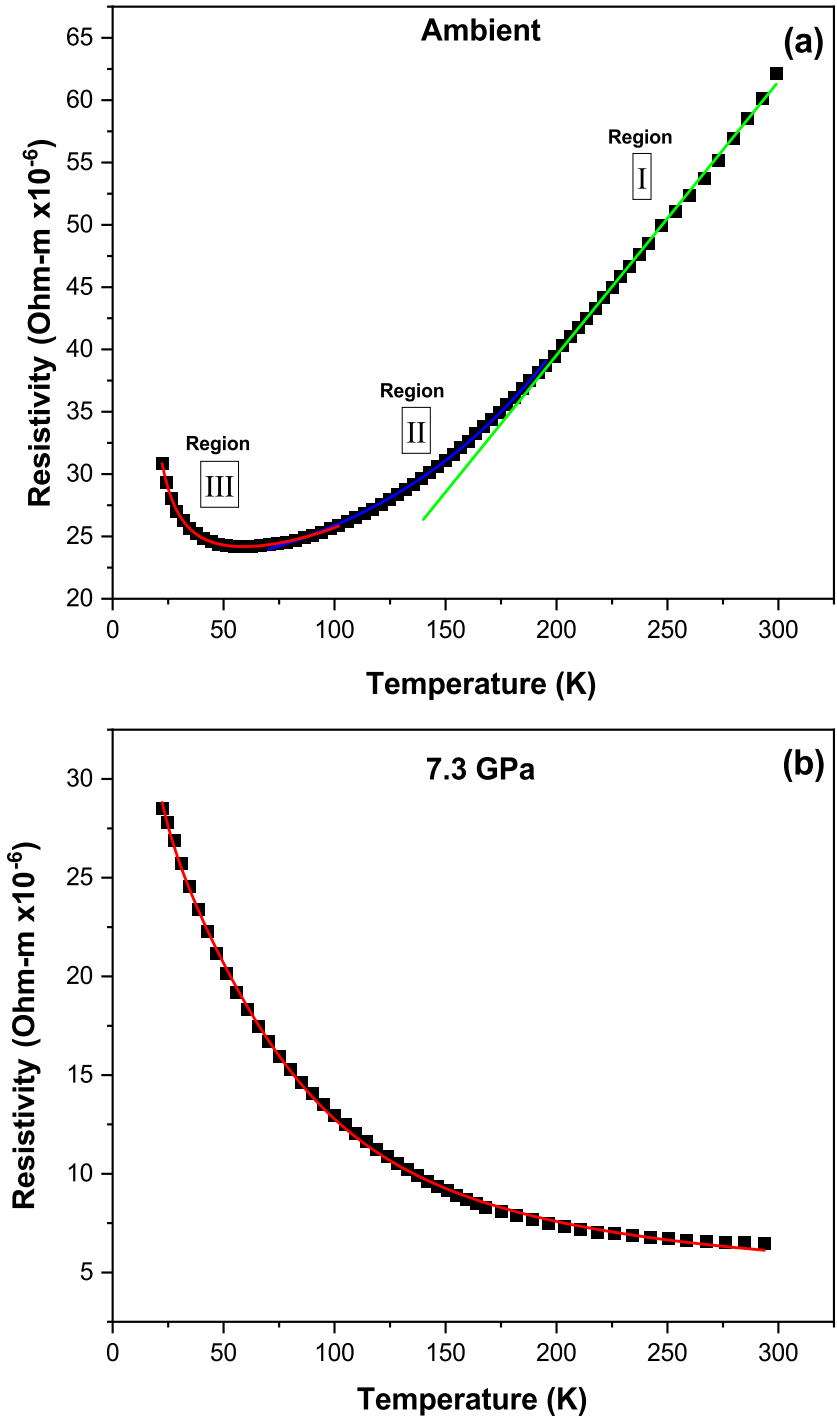}
	\caption{\label{Fig.7}(a)Temperature variation of resistivity at ambient pressure. Region I is fitted to Eq.(2), region II is fitted to Eq.(3) and region III is fitted to Eq.(4). Solid lines through the data points show respective fits. (b) Temperature variation of resistivity at 7.3 GPa, showing a semiconducting characteristic behaviour. The solid line through the data points shows the fit to the combined Eq.(4) and Eq.(5).}
\end{figure}

\end{document}


\title{Supplementary: Structural and electronic phase transitions in Zr$_{1.03}$Se$_{2}$ at high pressure}
	
	\author{Bishnupada Ghosh}
	\affiliation{National Center for High Pressure Studies, Department of Physical Sciences, Indian Institute of Science Education and Research Kolkata, Mohanpur 741246, Nadia, West Bengal, India.}
	
	\author{Mrinmay Sahu}
	\affiliation{National Center for High Pressure Studies, Department of Physical Sciences, Indian Institute of Science Education and Research Kolkata, Mohanpur 741246, Nadia, West Bengal, India.}
	
	\author{Debabrata Samanta}
	\affiliation{National Center for High Pressure Studies, Department of Physical Sciences, Indian Institute of Science Education and Research Kolkata, Mohanpur 741246, Nadia, West Bengal, India.}
	
	\author{Pinku Saha}
	\altaffiliation[Present address:]{Department of Earth Sciences, ETH Zürich, Zurich 8092, Switzerland}
	\affiliation{National Center for High Pressure Studies, Department of Physical Sciences, Indian Institute of Science Education and Research Kolkata, Mohanpur 741246, Nadia, West Bengal, India.}
	
	\author{Anshuman Mondal}
	\altaffiliation[Present address:]{Institut für Mineralogie, University of Münster, Münster 48149, Germany}
	\affiliation{National Center for High Pressure Studies, Department of Physical Sciences, Indian Institute of Science Education and Research Kolkata, Mohanpur 741246, Nadia, West Bengal, India.}

	\author{Goutam Dev Mukherjee}
	\affiliation{National Center for High Pressure Studies, Department of Physical Sciences, Indian Institute of Science Education and Research Kolkata, Mohanpur 741246, Nadia, West Bengal, India.}
	
	\email [Corresponding author:]{goutamdev@iiserkol.ac.in}

\date{\today}
\maketitle
\begin{figure}
	\includegraphics[width=\columnwidth]{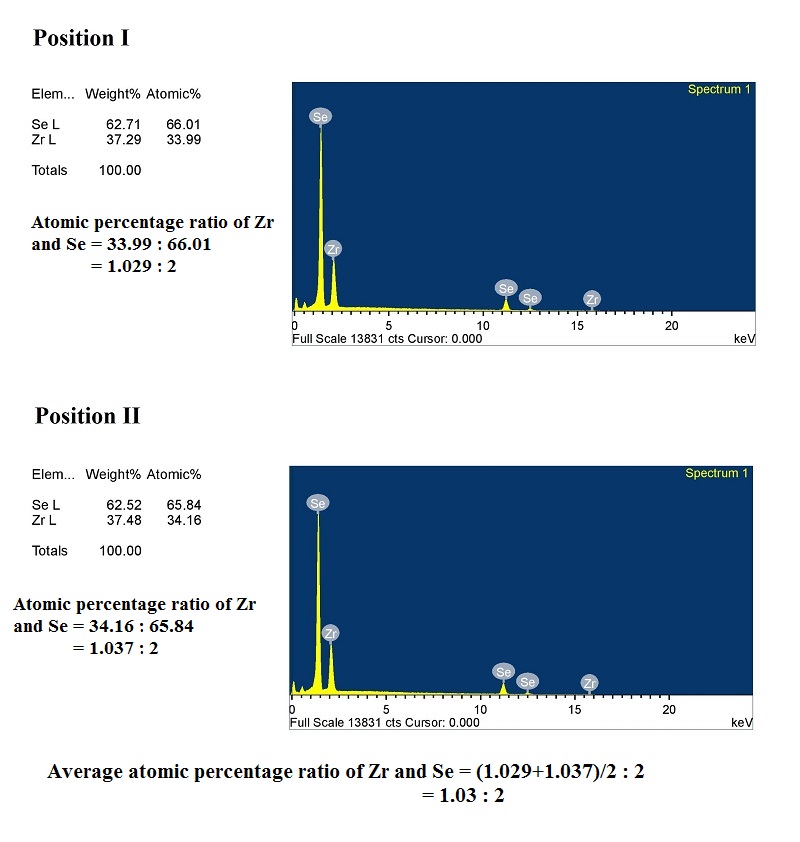}
	\caption{\label{Fig.1} Stoichiometric ratio of ZrSe$_{2}$ found using EDAX measurement at different positions.} 
\end{figure}

\begin{figure}
	\centering
	\includegraphics[width=\columnwidth]{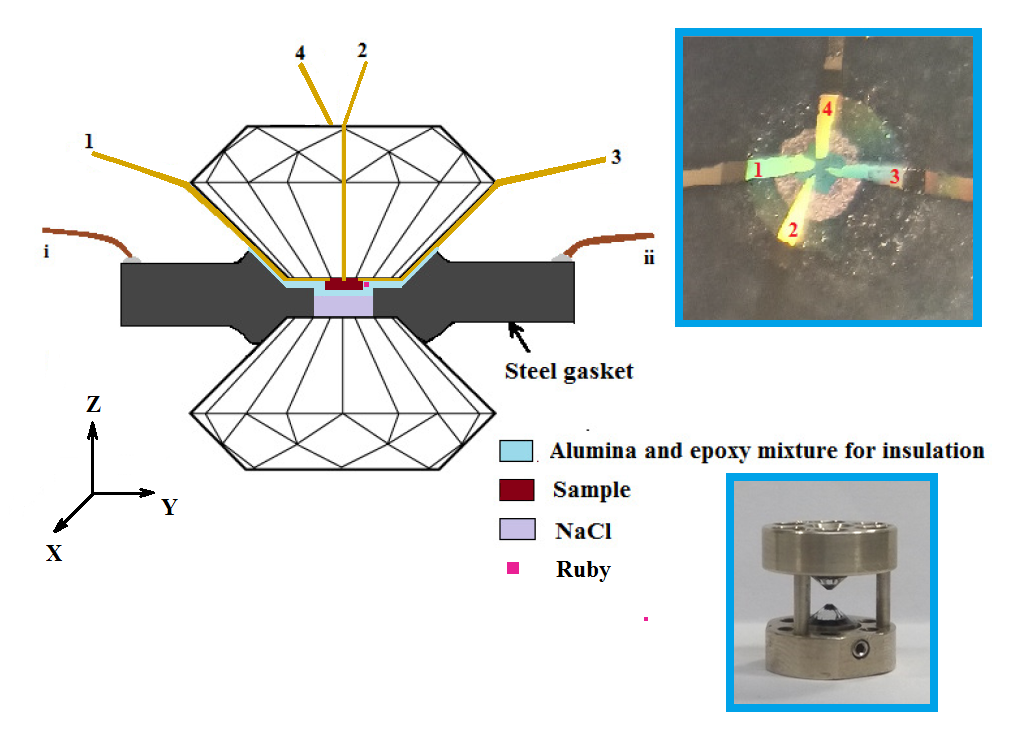}
	\caption{\label{Fig.2} Schematic diagram of loaded DAC for low temperature high pressure resistivity measurements. Used diamond anvil cell and sample with four lids are shown in insets.} 
\end{figure}
\begin{figure}
		\centering
		\includegraphics[width=\columnwidth]{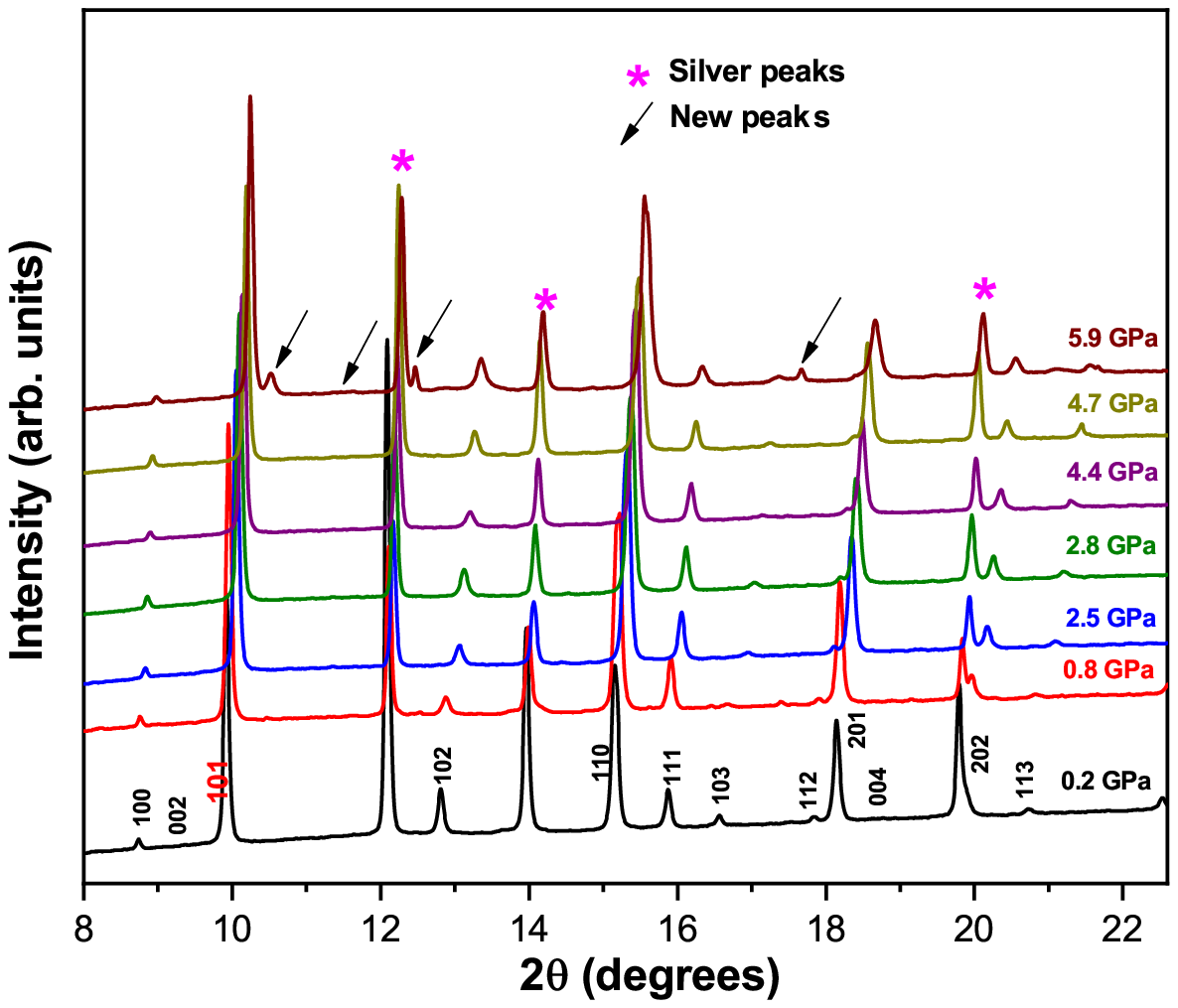}
		\caption{\label{Fig.3} Pressure evolution of XRD patterns at lower pressure region up to 5.9 GPa. New peaks are denoted by arrow and silver peaks are marked by star.} 
	\end{figure}

\begin{figure}
	\includegraphics[width=\columnwidth]{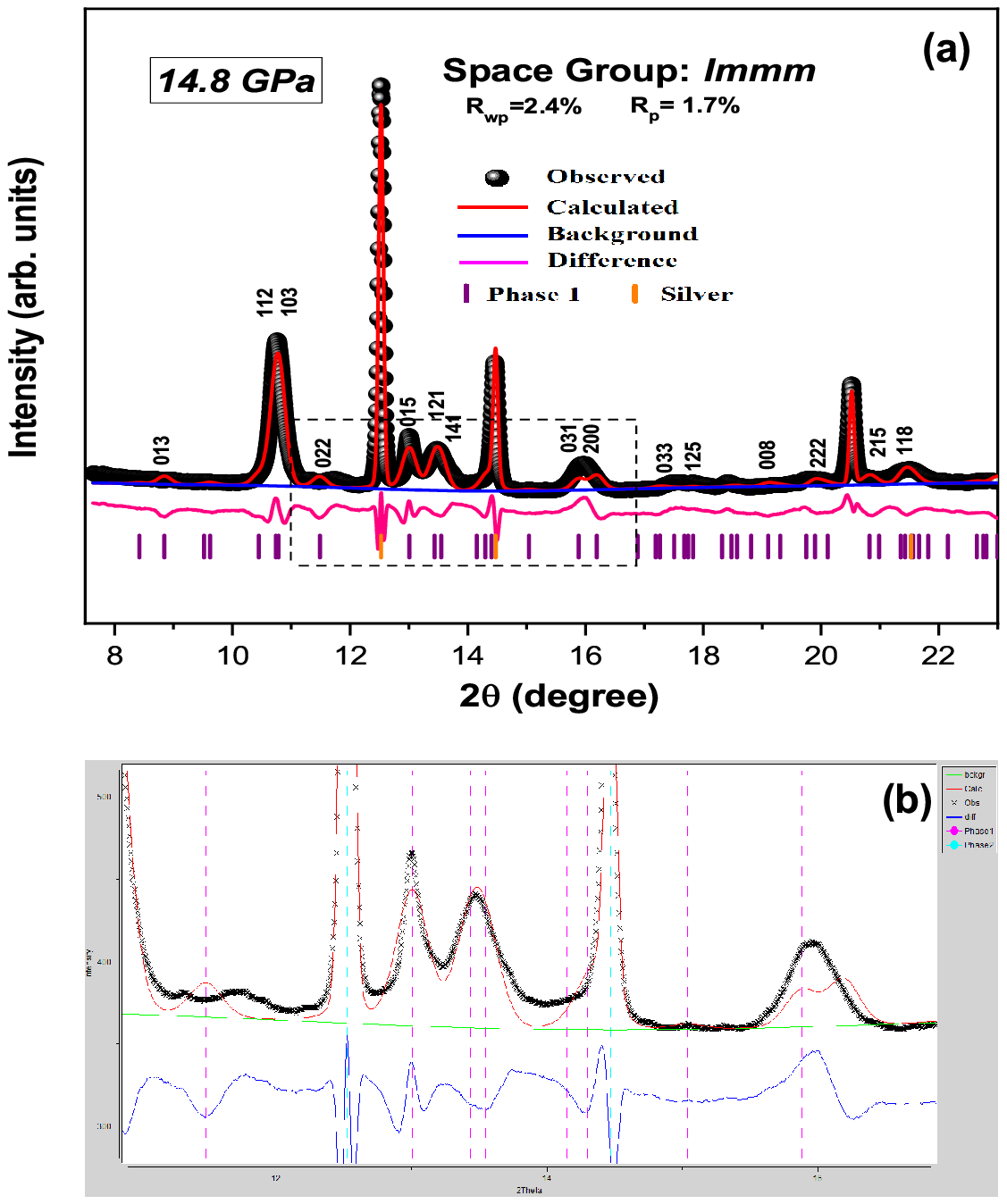}
	\caption{\label{Fig.4} (a) Rietveld refinement of XRD pattern at 14.8 GPa in the orthorhombic phase, (b) the part marked by box in (a) is zoomed and shown.} 
\end{figure}

\begin{figure}
	\centering
	\includegraphics[width=\columnwidth]{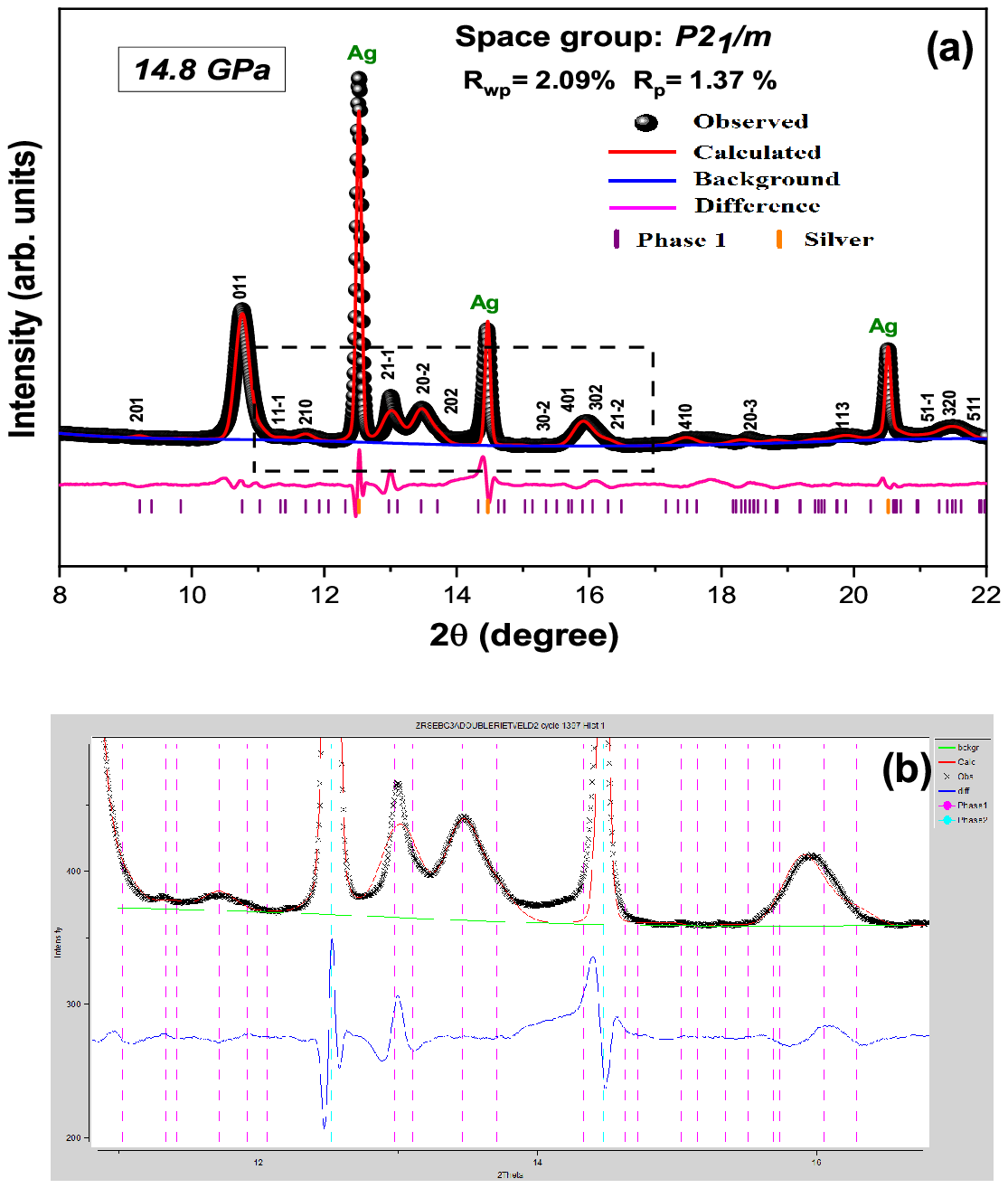}
	\caption{\label{Fig.5} (a) Rietveld refinement of XRD pattern at 14.8 GPa in the monoclinic phase, (b) the part marked by box in (a) is zoomed and shown.} 
\end{figure}

\begin{figure}
	\includegraphics[width=\columnwidth]{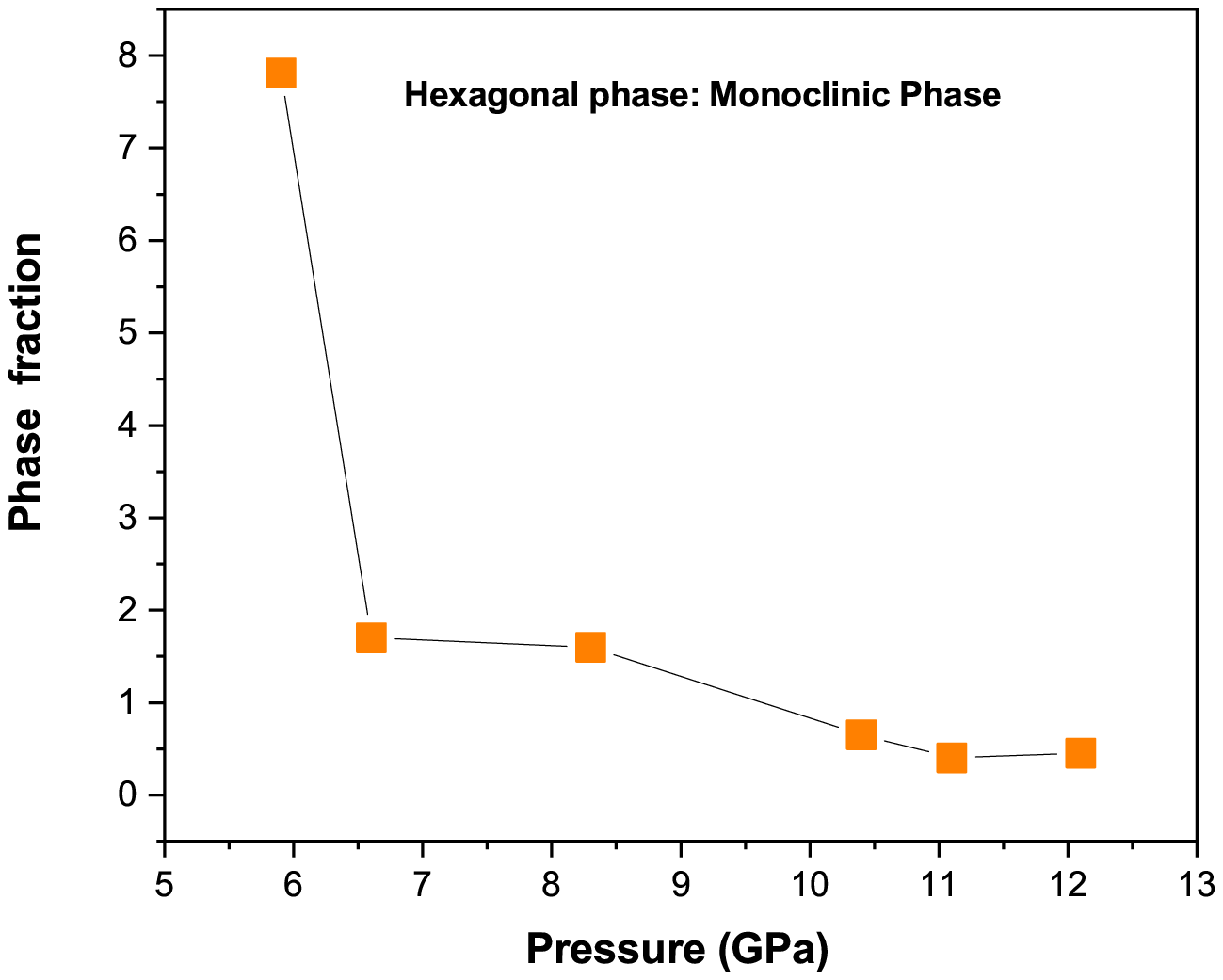}
	\caption{\label{Fig.6} Pressure variation of phase fraction of hexagonal and monoclinic phase.} 
\end{figure}

\begin{figure}
	\includegraphics[width=\columnwidth]{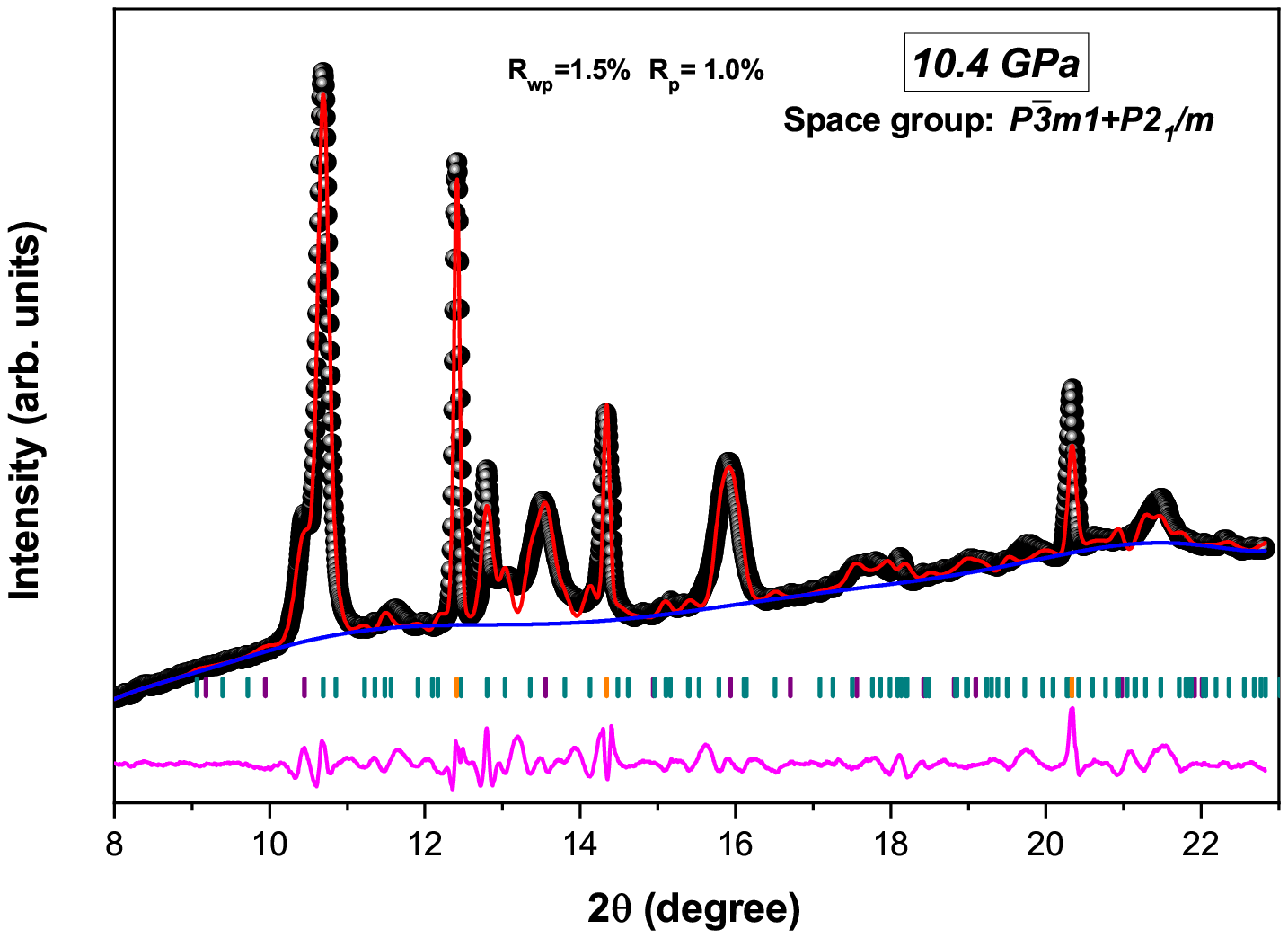}
	\caption{\label{Fig.7}Rietveld refinement of XRD pattern at 10.4 GPa in the mixed phase of hexagonal and monoclinic structure. } 
\end{figure}

\begin{figure}
	\includegraphics[width=\columnwidth]{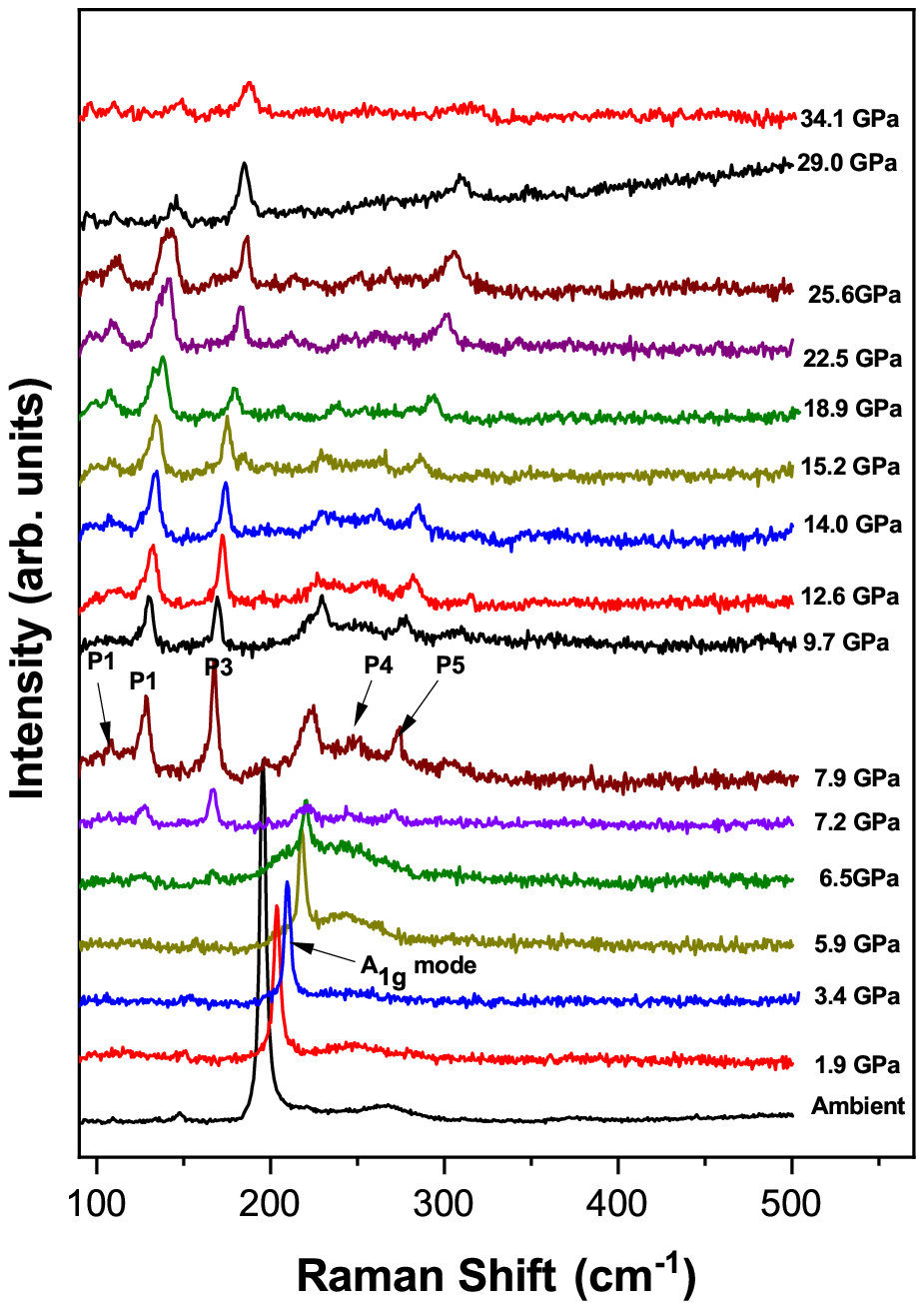}
	\caption{\label{Fig.8} Pressure evolution of Raman spectra up to 34.1 GPa.} 
\end{figure}

\begin{figure}
	\includegraphics[width=\columnwidth]{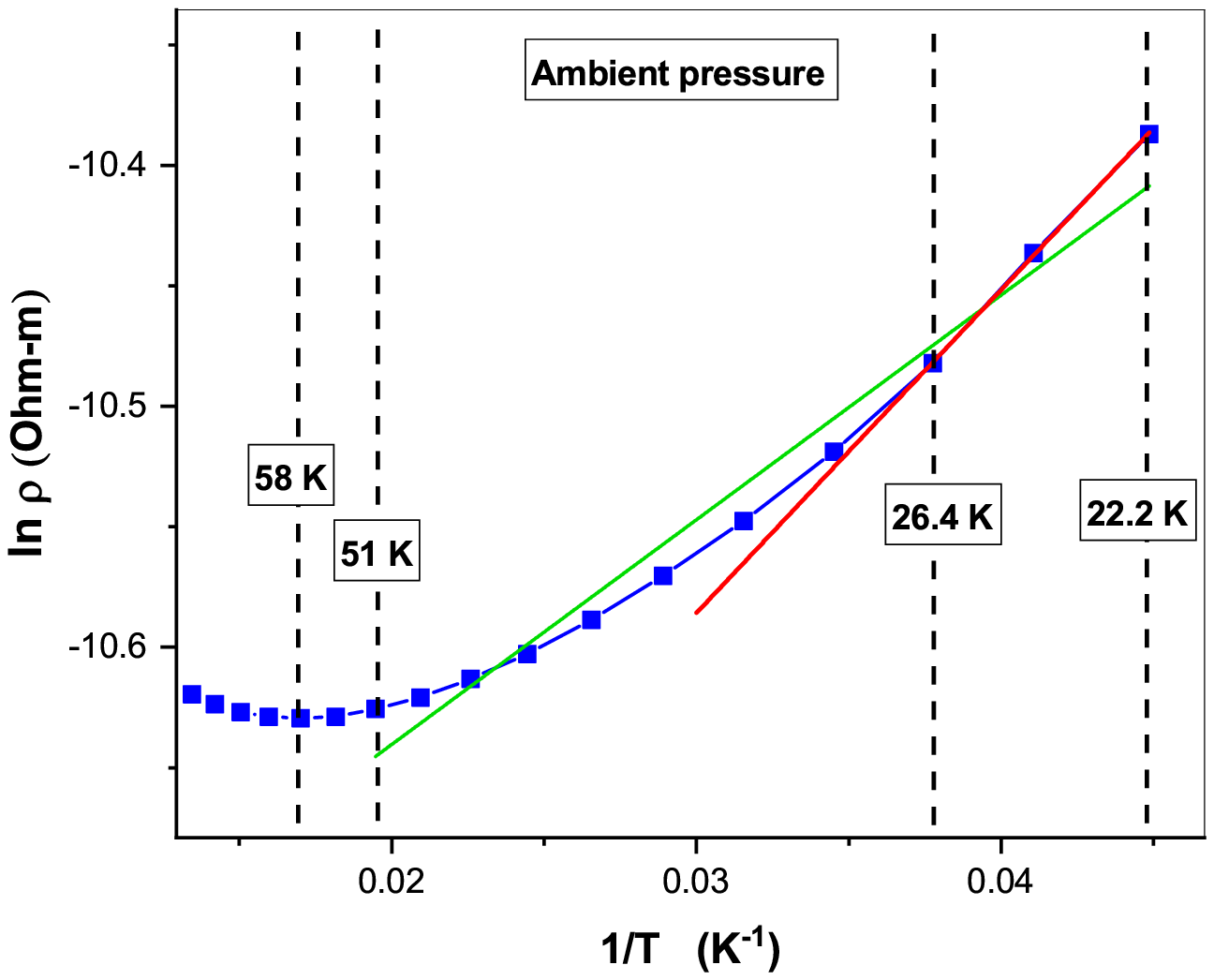}
	\caption{\label{Fig.9} ln$\rho$ vs 1/T plot at ambient pressure. The resistivity upturn at lower temperature region doesn't give a linear fit.} 
\end{figure}

\begin{figure}
	\includegraphics[width=\columnwidth]{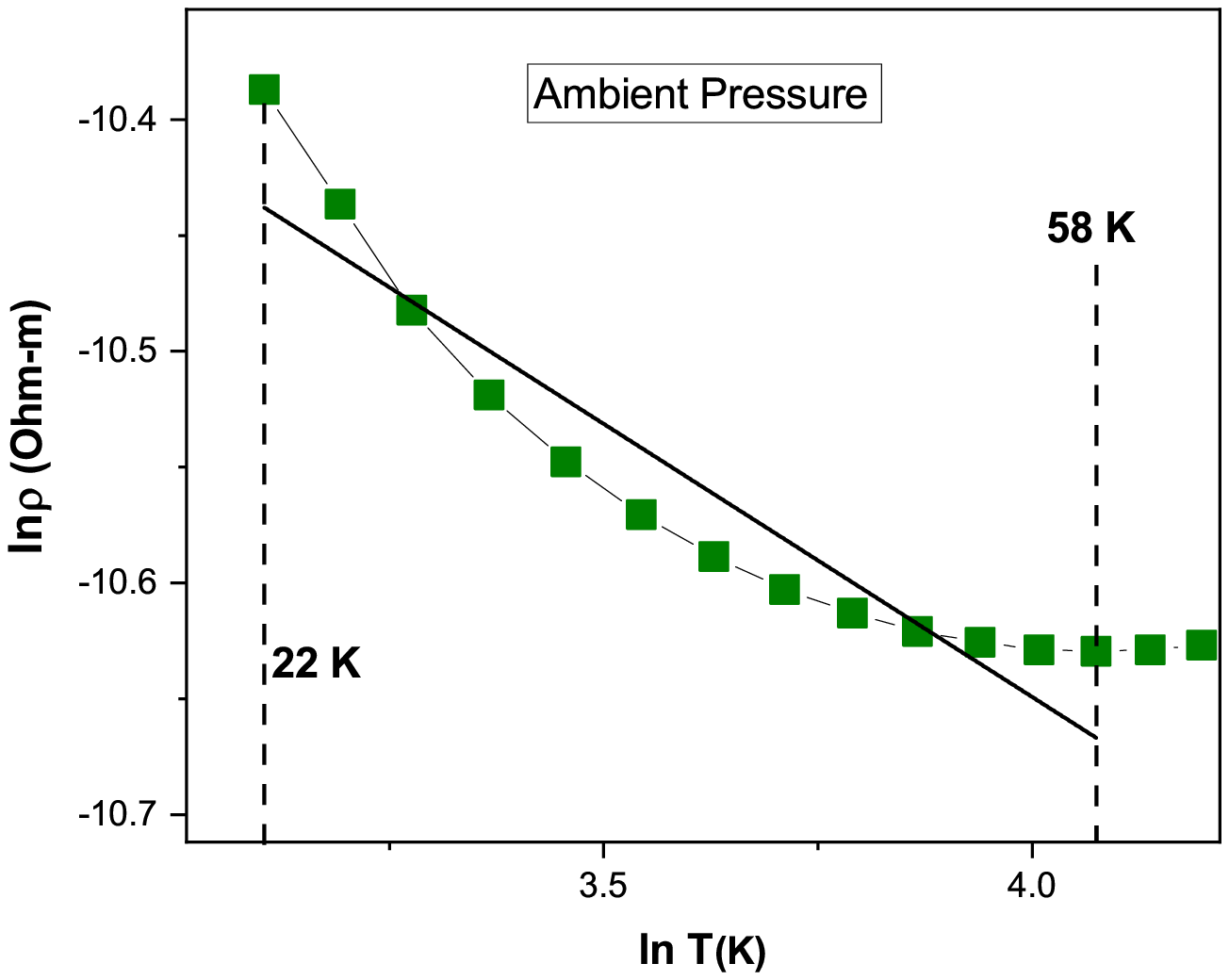}
	\caption{\label{Fig.10} ln$\rho$ vs lnT plot at ambient pressure. The resistivity upturn at lower temperature region doesn't give a linear fit.} 
\end{figure}
\begin{figure}
	\includegraphics[width=\columnwidth]{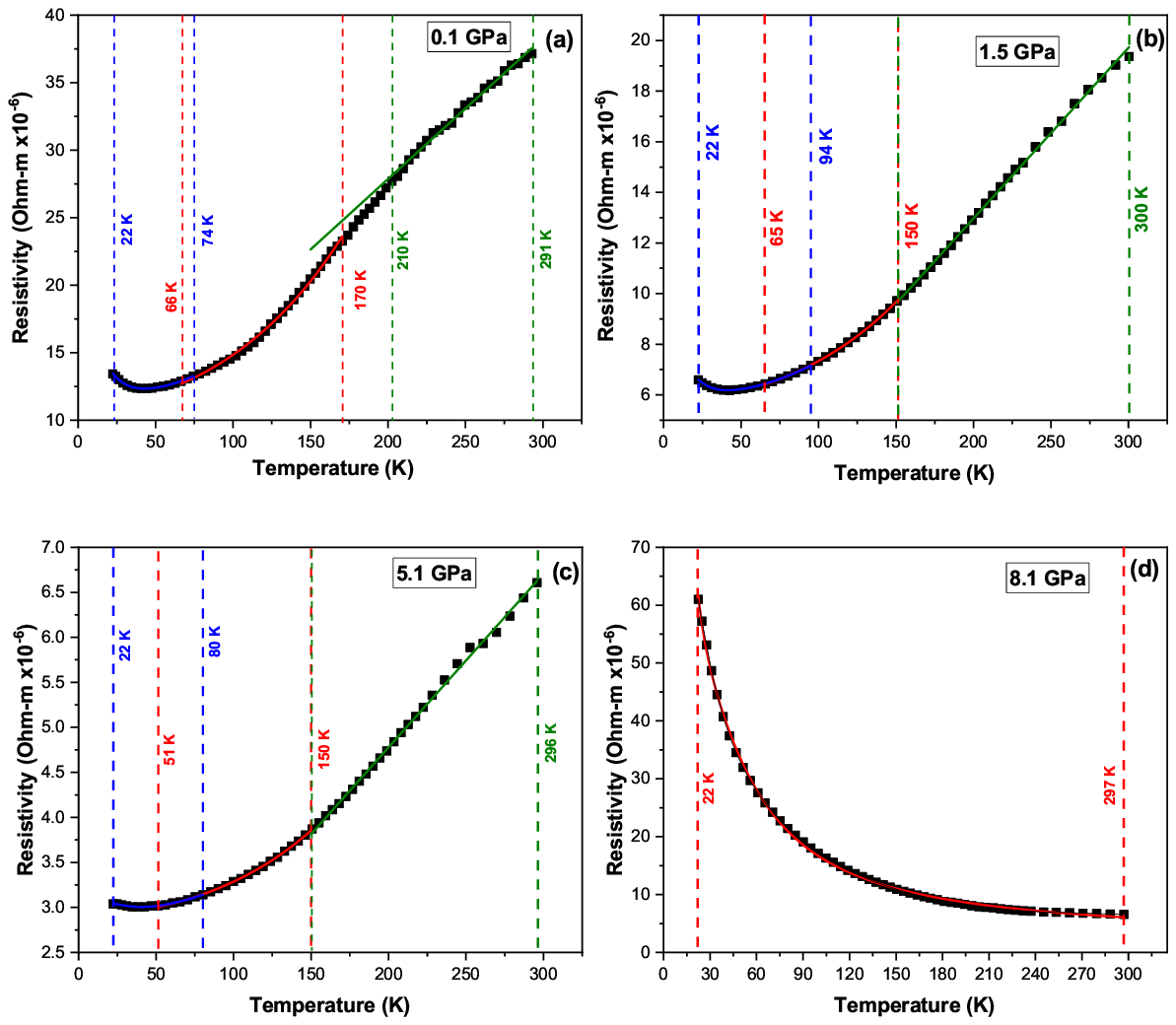}
	\caption{\label{Fig.11} Fitting of resistivity vs temperature data at different temperature regions at different pressure points.} 
\end{figure}
\begin{figure}
	\includegraphics[width=\columnwidth]{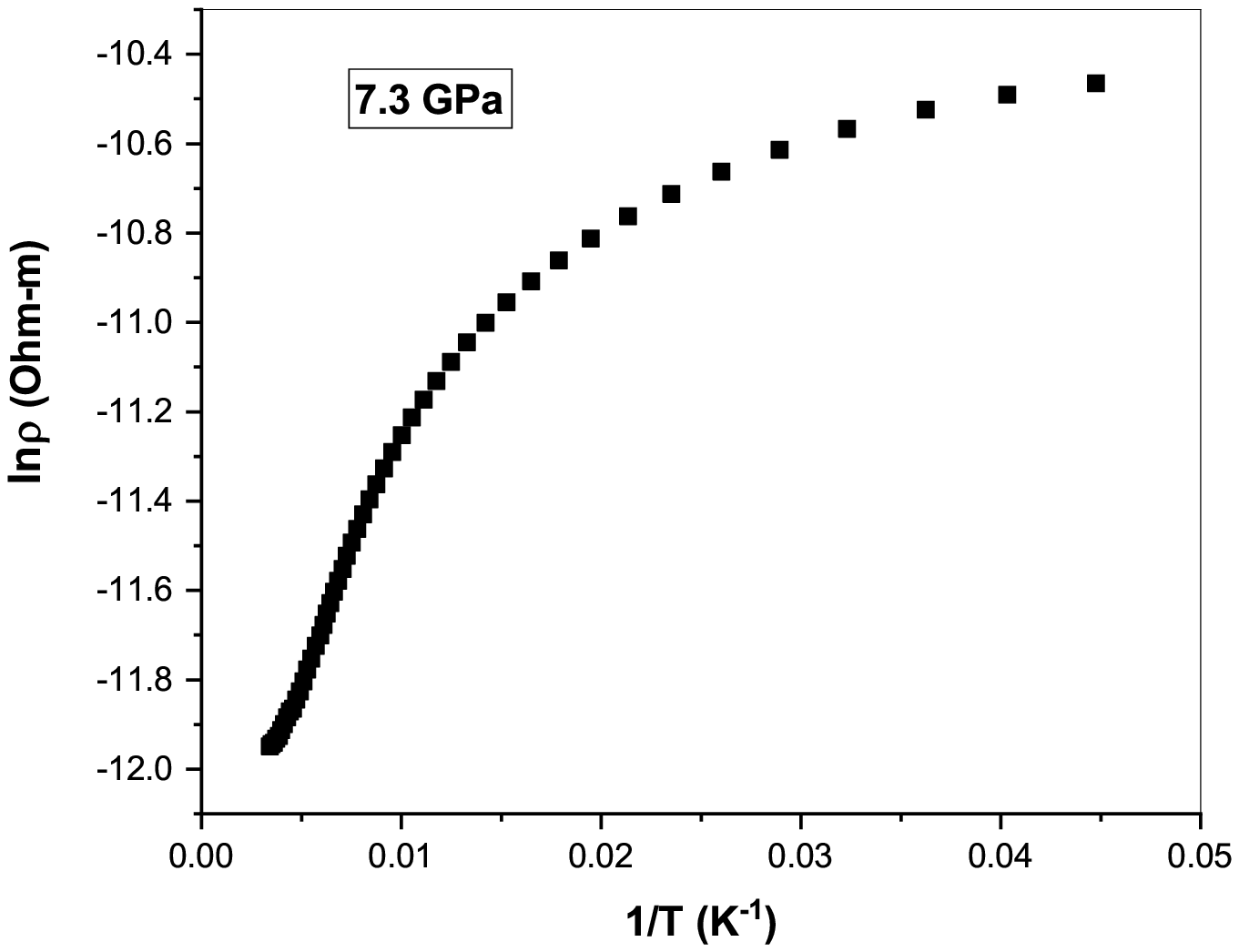}
	\caption{\label{Fig.12} ln$\rho$ vs 1/T plot at 7.3 GPa. The plot doesn't show any linear region.} 
\end{figure}
\begin{figure}
	\includegraphics[width=\columnwidth]{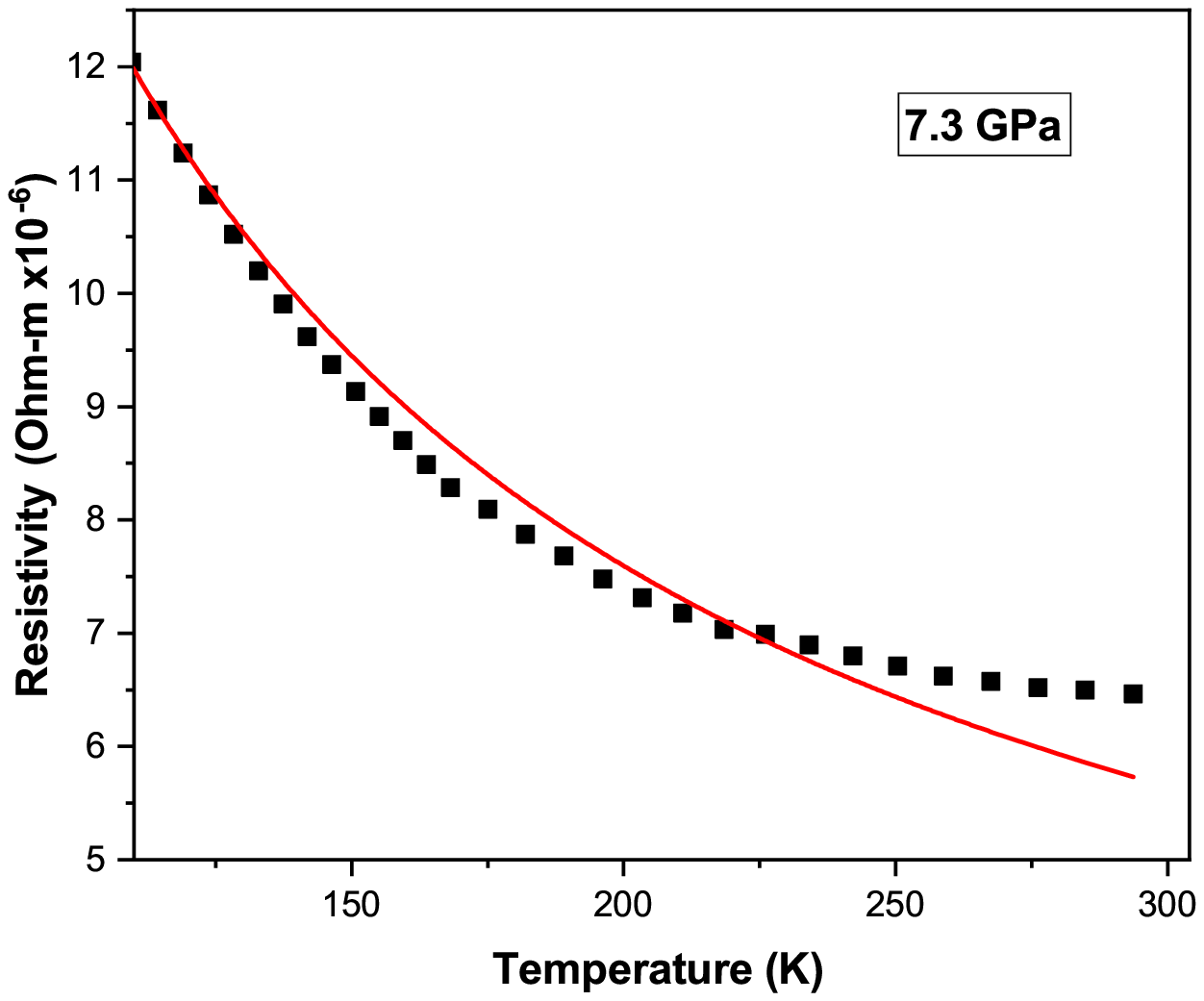}
	\caption{\label{Fig.13} The resistivity data at 7.3 GPa doesn't fit to logarithmic term due to the magnetic impurities. The fitted curve is shown using red line.} 
\end{figure}